\documentclass{elsart}
\usepackage{ragged2e}
\usepackage{graphics}
\usepackage{graphicx}
\usepackage{epsfig}
\usepackage{amssymb}
\usepackage{lineno}
\usepackage{blindtext}
\usepackage{lineno}
\usepackage{lineno, blindtext}
\usepackage{mathtools}
\begin{document}
	\begin{frontmatter}
\title{On the threshold of 1/2 order subharmonic emissions in the oscillations of ultrasonically excited bubbles}
\author{A.J. Sojahrood \thanksref{a},\thanksref{b}} \footnote{Email: amin.jafarisojahrood@ryerson.ca},
\author{H. Haghi \thanksref{a},\thanksref{b}}
\author{N.R. Shirazi}
\author{R. Karshafian \thanksref{a},\thanksref{b}}
\author{M.C. Kolios \thanksref{a},\thanksref{b}}
\address[a]{Department of Physics, Ryerson University, Toronto, Canada}
\address[b]{Institute for Biomedical Engineering, Science and Technology (IBEST) a partnership between Ryerson University and St. Michael's Hospital, Toronto, Ontario, Canada}
\begin{abstract}
	The pressure threshold for 1/2 order subharmonic (SH) emissions and period doubling during the oscillations of ultrasonically excited bubbles is thought to be minimum when the bubble is sonicated with twice its resonance frequency ($f_r$). This estimate is based on studies that simplified or neglected the effects of thermal damping. In this work, the nonlinear dynamics of ultrasonically excited bubbles is investigated accounting for the thermal dissipation. Results are visualized using bifurcation diagrams as a function of pressure. Here we show that, and depending on the gas, the pressure threshold for 1/2 order SHs can be minimum at a frequency between $0.5f_r\leq f \leq0.6f_r$. In this frequency range, the generation of 1/2 order SHs are due to the occurrence of 5/2 order ultra-harmonic resonance. The stability of such oscillations are size dependent. For an air bubble immersed in water, only bubbles bigger than 1 $\mu m$ in diameter are able to emit non-destructive SHs in these frequency ranges.   
\end{abstract}
\end{frontmatter}

\section{Introduction}
An ultrasonically excited bubble is a nonlinear oscillator that is present in a wide range of applications and phenomena from underwater acoustics to material science and medicine \cite{1}.
 Starting with the pioneering work of Parlitz et al \cite{2}, the bifurcation structure of the bubble oscillators exhibiting period doubling and chaos has been the topic of many recent investigations \cite{3,4,5,6,7}. Period doubling (PD)  results in the emission of 1/2 order SHs and (3/2,5/2, etc) UHs by bubbles. These nonlinear oscillators have many applications in diagnostic and therapeutic ultrasound \cite{8,9,10,11,12,13,14} including the SH and UH emissions by bubbles is in monitoring the blood brain barrier opening by ultrasound \cite{15,16}. \\  
Cavitation and bubble oscillations are generally classified into two types, stable cavitation, which results in emissions at subharmonics of the main excitation frequency \cite{16,17} and inertial cavitation, which is characterized by broadband noise emissions \cite{17}. Understanding the generation and amplification of SHs as well as differentiating between stable and inertial regimes are critical for several applications \cite{17,18,19,20,21,22,23}.  This is because gentle bubble oscillations may result in more control over the bubble dynamics and less unwanted tissue damage. In non-medical applications like material cleaning, stable non-destructive bubble oscillations are an additional mechanism of surface cleaning thus eliminating erosive bubble collapse \cite{24,25}. Potential underlying mechanisms of cleaning has been investigated in \cite{25} and one of the cleaning mechanisms is through steadily fatiguing the surface contamination by the non-destructive stable micro-streaming around bubbles.\\ Among many causes, SH emissions can be generated due to nonlinear volumetric oscillations of the bubbles \cite{26,27,28,29}, from the onset of surface modes \cite{30,31} or from energy transfer from surface to volume oscillations \cite{32}.\\ Many studies have investigated the pressure threshold of SH emissions by ultrasonically excited bubbles \cite{29,33,34,35,36,37,38,39}. These studies indicate that, the pressure threshold for SH emissions is the lowest when the bubble is sonicated with a frequency that is twice its resonance frequency ($f_r$) \cite{29,33,34,35,36,37,38,39}. More detailed numerical investigations in \cite{37,38} revealed that the pressure threshold for SH oscillations shift to frequencies near resonance for bubbles smaller than $0.6 \mu m$ in diameter. The mechanism behind the shift in the minimum pressure threshold is attributed to the increased damping due to viscosity on smaller bubbles. Katiyar and Sarkar \cite{38,39} conclude that the increased dissipation due to viscosity damps the bubble response more at twice the resonance than at resonance, leading to a shift of the minimum from twice the resonance frequency toward the resonance frequency. The addition of shell has been shown analytically and numerically in \cite{38,39} to decrease the pressure threshold for SH generation and possibly shift the frequency of minimum pressure threshold towards resonance. The impact of bubble-bubble interaction on the threshold of SH emissions has been investigated by Guerda et al \cite{40}. They show that, for a given bubble radius, increasing the concentration of bubbles shifts the subharmonic resonance frequency towards lower values and reduces the
minimum pressure threshold for SH generation.\\ 
The previous studies either simplified the thermal damping using the linear models \cite{41,42} or ignored thermal effects. To provide a more accurate understanding of the effect of dissipation on the SH emissions by bubbles, we have considered the effect of thermal damping using the full set of ordinary differential equations (ODEs) that describe the thermal effects \cite{43,44}. Using the more accurate description of thermal effects, the bifurcation structure of the radial oscillations of the bubbles is studied as a function of pressure for driving frequencies between $0.1f_r<f<2.2f_r$. We used a more comprehensive method for bifurcation analysis introduced in \cite{45}. We show that contrary to the common belief and depending on the gas, insonation with frequencies between $0.3f_r<f<0.7f_r$ leads to the minimum pressure threshold for 1/2 order SH emissions. A more careful analysis of the oscillations reveals that 3/2 or 5/2 ultraharmonic (UH) resonance is the mechanism behind the generation of SHs in this frequency range.

\section{Methods}
\subsection{The bubble model}
The dynamics of the uncoated bubble including the compressibility effects to the first order of Mach number can be modeled using Keller-Miksis (KM) equation\cite{46}:
\justifying
\begin{equation}
\rho[(1-\frac{\dot{R}}{c})R\ddot{R}+\frac{3}{2}\dot{R}^2(1-\frac{\dot{R}}{3c})]=\\
(1+\frac{\dot{R}}{c}+\frac{R}{c}\frac{d}{dt})(P_g-\frac{4\mu_L\dot{R}}{R}-\frac{2\sigma}{R}-P_0-P_A sin(2 \pi f t))
\end{equation}
In this equation, R is radius at time t, $R_0$ is the initial bubble radius, $\dot{R}$ is the wall velocity of the bubble, $\ddot{R}$ is the wall acceleration,	$\rho{}$ is the liquid density (998 $\frac{kg}{m^3}$), c is the sound speed (1481 m/s), $P_g$ is the gas pressure.\\ In the absence of thermal effects $P_g$ is given by $P_g=(P_0+2\frac{\sigma}{R_0})(\frac{R_0}{R})^{3\gamma}$, $\sigma{}$ is the surface tension (0.0725 $\frac{N}{m}$), $\mu_L$ is the liquid viscosity (0.001 Pa.s), and $P_A$ and \textit{f} are the amplitude and frequency of the applied acoustic pressure. The values in the parentheses are for pure water at 293 K. In this paper the gas inside the uncoated bubble is either air or C3F8, and water is the host media. Thermal properties for the gases are given in Table 1. \\
\begin{table}
	\begin{tabular}{ |p{2cm}||p{4.5cm}|p{2cm}|p{2cm}|p{2cm}|  }
		\hline
		\multicolumn{5}{|c|}{Thermal parameters of the gases at 1 atm} \\
		\hline
		Gas type  & L($\frac{W}{mK}$) &$c_p$$(\frac{kJ}{kg K})$ &$c_v$ $(\frac{kJ}{kg K})$&$\rho_g$ $(\frac{kg}{m^3})$\\
		\hline
		Air \cite{47}   & 0.01165+$5.528\times10^{25}\times T^2$ &1.0049&   0.7187&1.025\\
		C3F8 \cite{48} &   0.012728  & 0.79   &0.7407&8.17\\
		\hline
	\end{tabular}
	\caption{Thermal properties of the gases used in simulations.}
	\label{table:1}
\end{table}
\subsection{Thermal effects}
\subsubsection{ODE thermal model}
If thermal effects are considered, $P_g$ is given by Eq. 2 \cite{43,44}:
\begin{equation}
P_g=\frac{N_gKT}{\frac{4}{3}\pi R(t)^3-N_g B}
\end{equation}
Where $N_g$ is the total number of the gas molecules, $K$ is the Boltzman constant and B is the molecular co-volume of the gas inside the bubble. The average temperature inside the gas can be calculated using Eq. 3 \cite{43,44}:
\begin{equation}
\dot{T}=\frac{4\pi R(t)^2}{C_v} \left(\frac{L\left(T_0-T\right)}{L_{th}}-\dot{R}P_g\right)
\end{equation}
where $C_v$ is the heat capacity at constant volume, $T_0$=$293 $ K is the initial gas temperature, $L_{th}$ is the thickness of the thermal boundary layer. $L_{th}$ is given by $L_{th}=min(\sqrt{\frac{DR(t)}{|\dot{R(t)}|}},\frac{R(t)}{\pi})$ where $D$ is the thermal diffusivity of the gas which can be calculated using $a=\frac{L}{c_p \rho_g}$ where L is the gas thermal conductivity and $c_p$ is specific heat capacity at constant pressure and $\rho_g$ is the gas density.\\
Predictions of the ODE thermal model have been shown to be in good agreement with predictions of the models that incorporate the thermal effects using the PDEs \cite{45} that incorporate the temperature gradients within the bubble.
To calculate the radial oscillations of the uncoated bubble while including the thermal effects, Eqs. 1 is coupled with Eq. 2 and 3 and is then solved using the ode45 solver of Matlab.
\subsubsection{Linear thermal model}
The linear thermal model \cite{41,42} is a popular model that has been widely used in studies related to oceanography \cite{49,50} and the modeling and characterization of coated bubble oscillations \cite{51,52,53,54,55}. In this model thermal damping is approximated through linearization by adding an artificial viscosity term to the liquid viscosity. Furthermore, a variable isoentropic index is used instead of the polytropic exponent of the gas.\\ In this model $P_g$ is given by:
\begin{equation}
P_g=P_{g0}\left(\frac{R_0}{R}\right)^{3k}
\end{equation} 
Where the polytropic exponent $\gamma$ is replaced by isoentropic indice ($k$):
\begin{equation}
k=\frac{1}{3}\Re(\phi)
\end{equation}
where $\Re$ denotes the real part.\\ 
The liquid viscosity is artificially increased by adding a thermal viscosity ($\mu_{th}$) to the liquid viscosity. The thermal viscosity ($\mu_{th}$) is given by:
\begin{equation}
\mu_{th}=\frac{P_{g0} \Im(\phi)}{\omega}
\end{equation}
In the above equations $\Im$ denotes the imaginary part and the complex term $\phi$ is calculated from
\begin{equation}
\phi=\frac{3\gamma}{1-3\left(\gamma-1\right)i\chi\left[\left(\frac{i}{\chi}\right)^{\frac{1}{2}}coth\left(\frac{i}{\chi}\right)^{\frac{1}{2}}-1\right]}
\end{equation}
where $\gamma$ is the polytropic exponent and $\chi=\frac{D}{\omega R_0^2}$ represents the thermal diffusion length. $D$ is the thermal diffusivity of the gas. $D= \frac{L}{c_p \rho_g}$ where $c_p$, $\rho_g$, and $L$ are the specific heat capacity at constant pressure, density and thermal conductivity of the gas inside the bubble.\\
To calculate the radial oscillations of the bubble while including the linear thermal effects, Eq. 1  is coupled with Eq. 4 and liquid viscosity is increased by $\mu_{th}$ (Eq. 6). 
\subsection{Investigation techniques}
Bifurcation diagrams are valuable tools to analyze the dynamics of nonlinear systems where the qualitative and quantitative changes of the dynamics of the system can be investigated effectively over a wide range of the control parameters. In this paper, we employ a more comprehensive bifurcation analysis method introduced in \cite{45}.
\subsubsection{Poincaré section}
When dealing with systems responding to a driving force, one option is to sample the R(t) curves using a specific point for each driving period. The approach can be summarized in:
\begin{equation}
P \equiv (R(\Theta))\{(R(t),  \dot{R}(t) ):\Theta= \frac{n}{f} \}\hspace{0.1cm} 
\end{equation}
where $P$ denotes the points in the bifurcation diagram, $R$ and $\dot{R}$
are the time dependent radius and wall velocity of the bubble at a given
set of control parameters of ($R_{0}$, $P_{A}$ and $f$), $\Theta$ is given by $\frac{n}{f}$ and n=1,2,....440. Points on the bifurcation diagram are constructed by plotting the solution of $R(t)$ at time points that are multiples of the driving acoustic period. The results are plotted for $n=400-440$ to ensure a steady state solution has been reached for all bubbles.
\subsubsection{Method of peaks}
Another way of constructing bifurcation points is by setting one of the phase space coordinates to zero:      
\begin{equation}
Q \equiv max(R)\{(R, \dot{R} ):\dot{R}= 0\}
\end{equation}
In this method, the steady state solution of the radial oscillations for each control parameter is considered. The maxima of the radial peaks ($\dot{R}=0$ and $\ddot{R}>0$) are identified (determined within 400-440 cycles of the stable oscillations) and are plotted versus the given control parameter in the bifurcation diagrams.\\ 
The bifurcation diagrams of the normalized bubble oscillations ($R/R_0$) are calculated using both methods a) and b). When the two results are plotted alongside each other, it is easier to uncover more important details about the SuH and UH oscillations, as well as the SH and chaotic oscillations. 

\begin{figure*}
	\begin{center}
		\scalebox{0.3}{\includegraphics{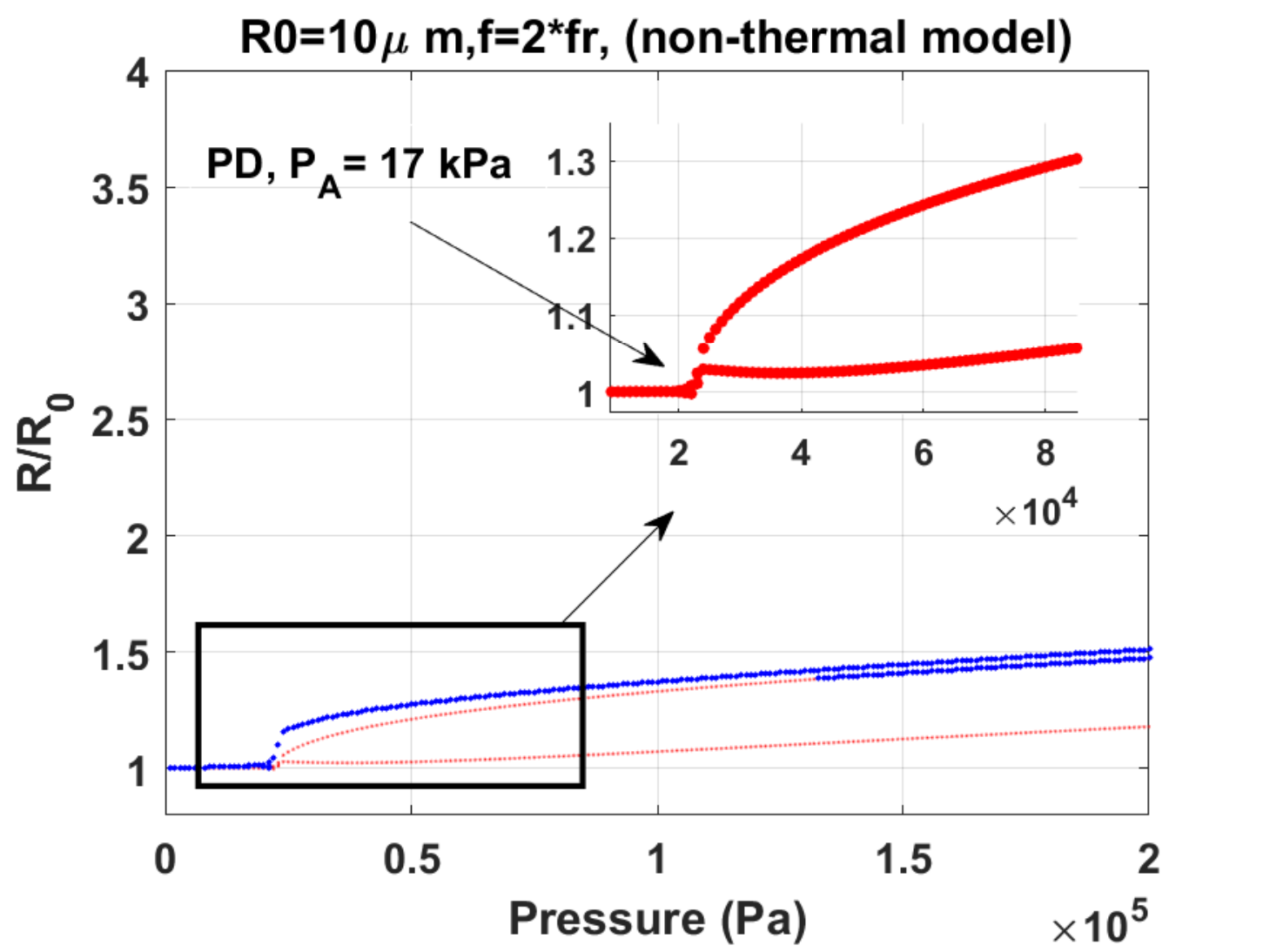}}\scalebox{0.3}{\includegraphics{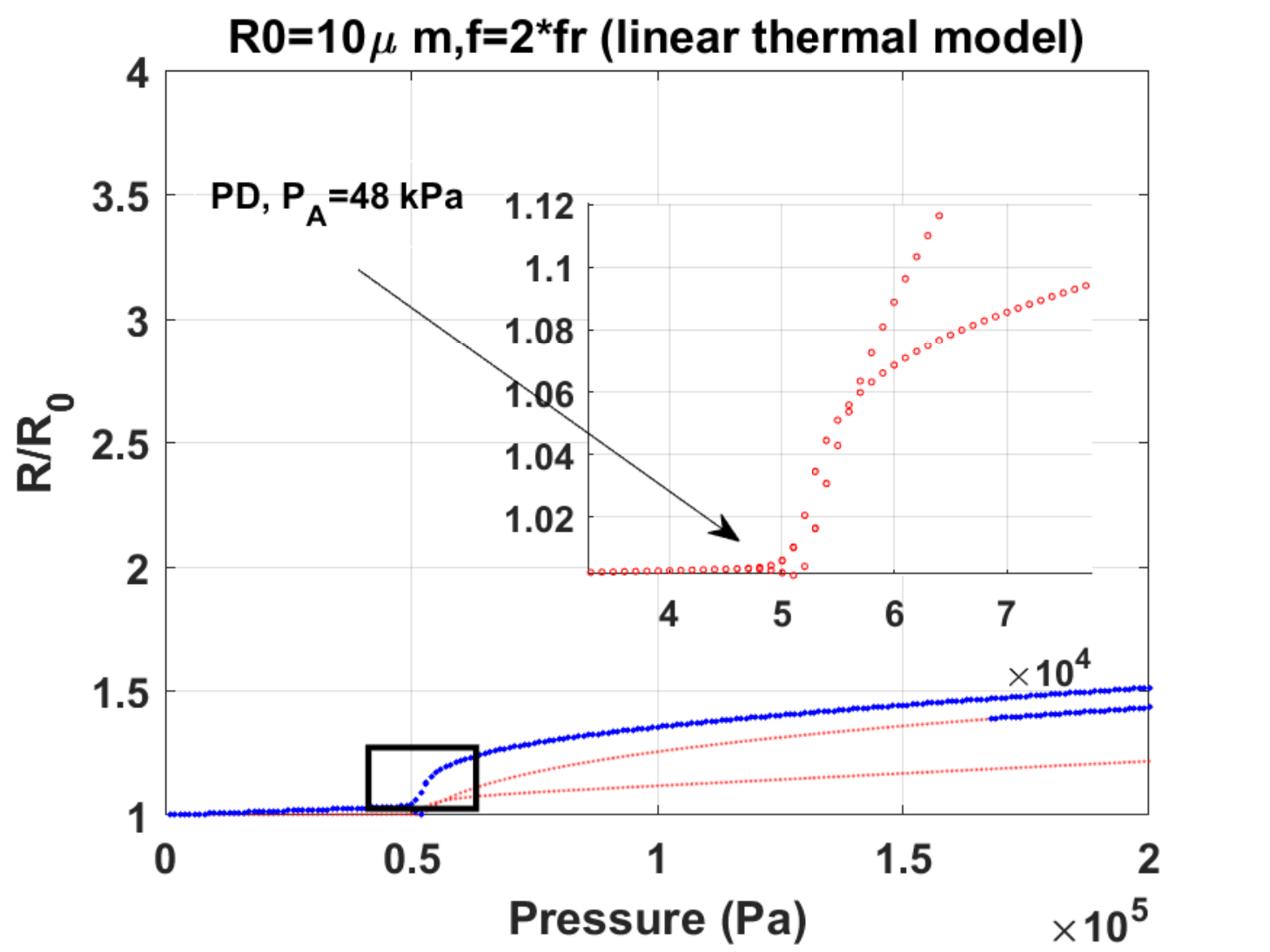}}	\scalebox{0.3}{\includegraphics{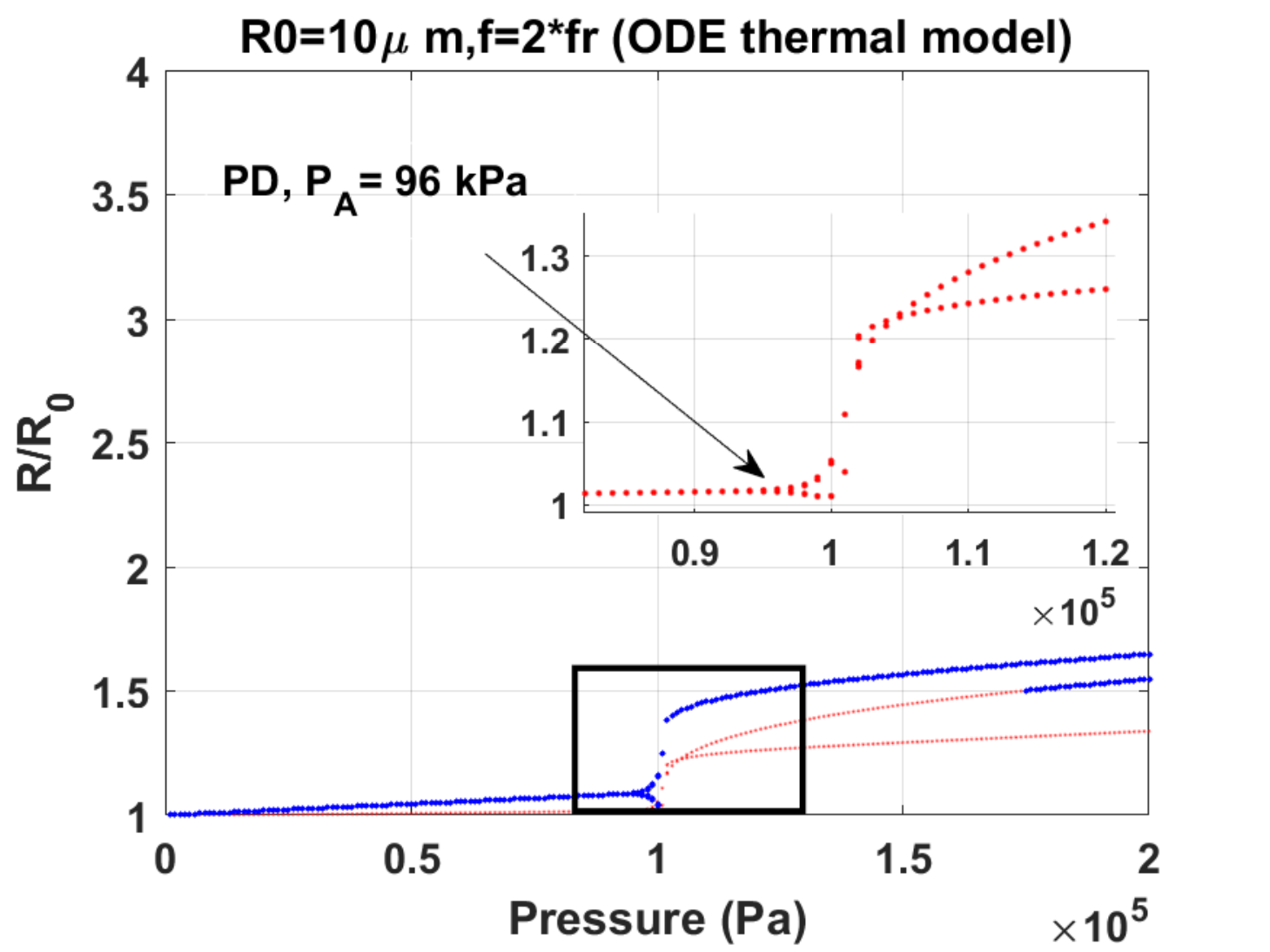}}\\
		(a) \hspace{3.5cm} (b) \hspace{3.5cm} (c) \\
		\scalebox{0.3}{\includegraphics{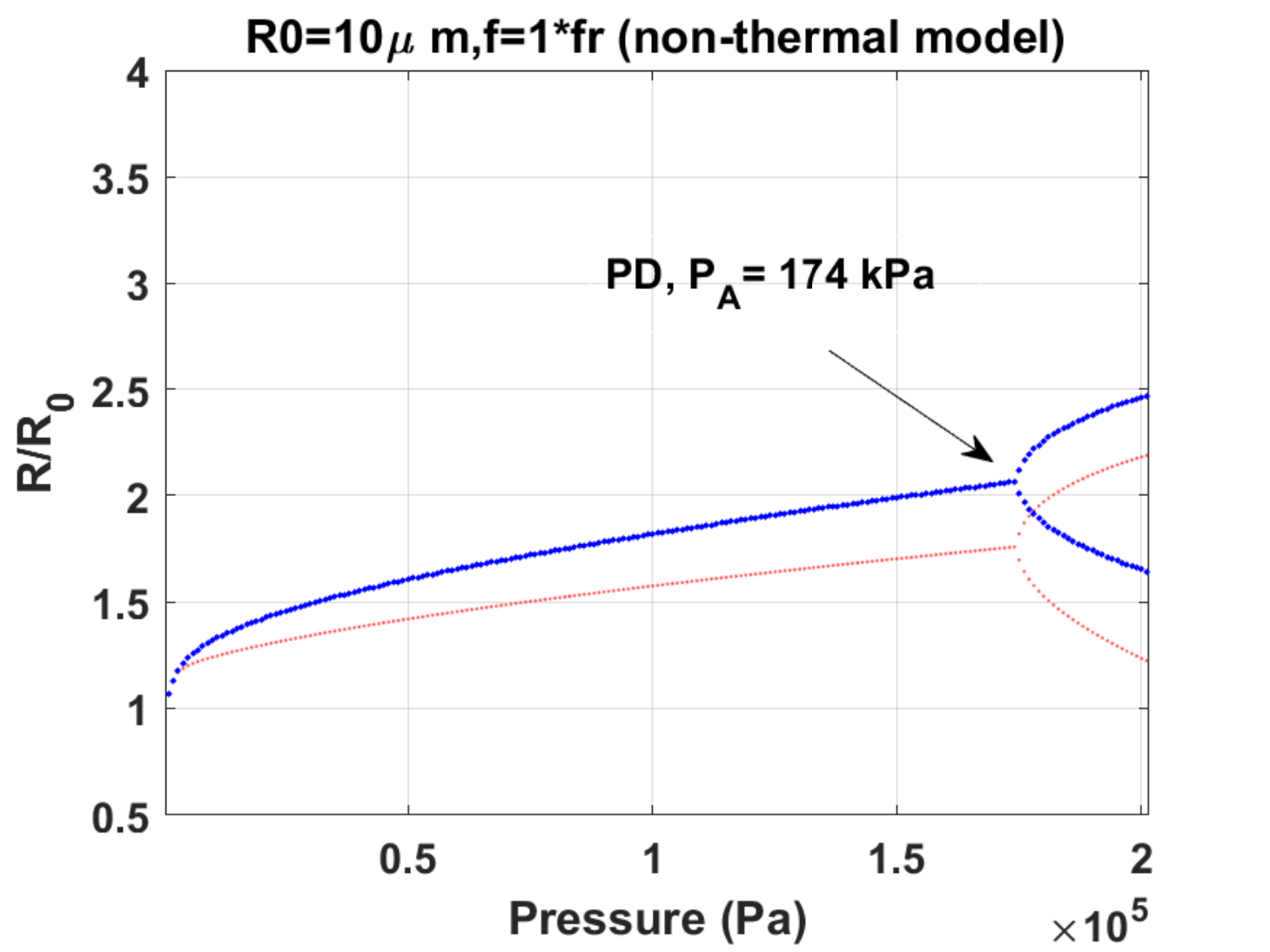}}\scalebox{0.3}{\includegraphics{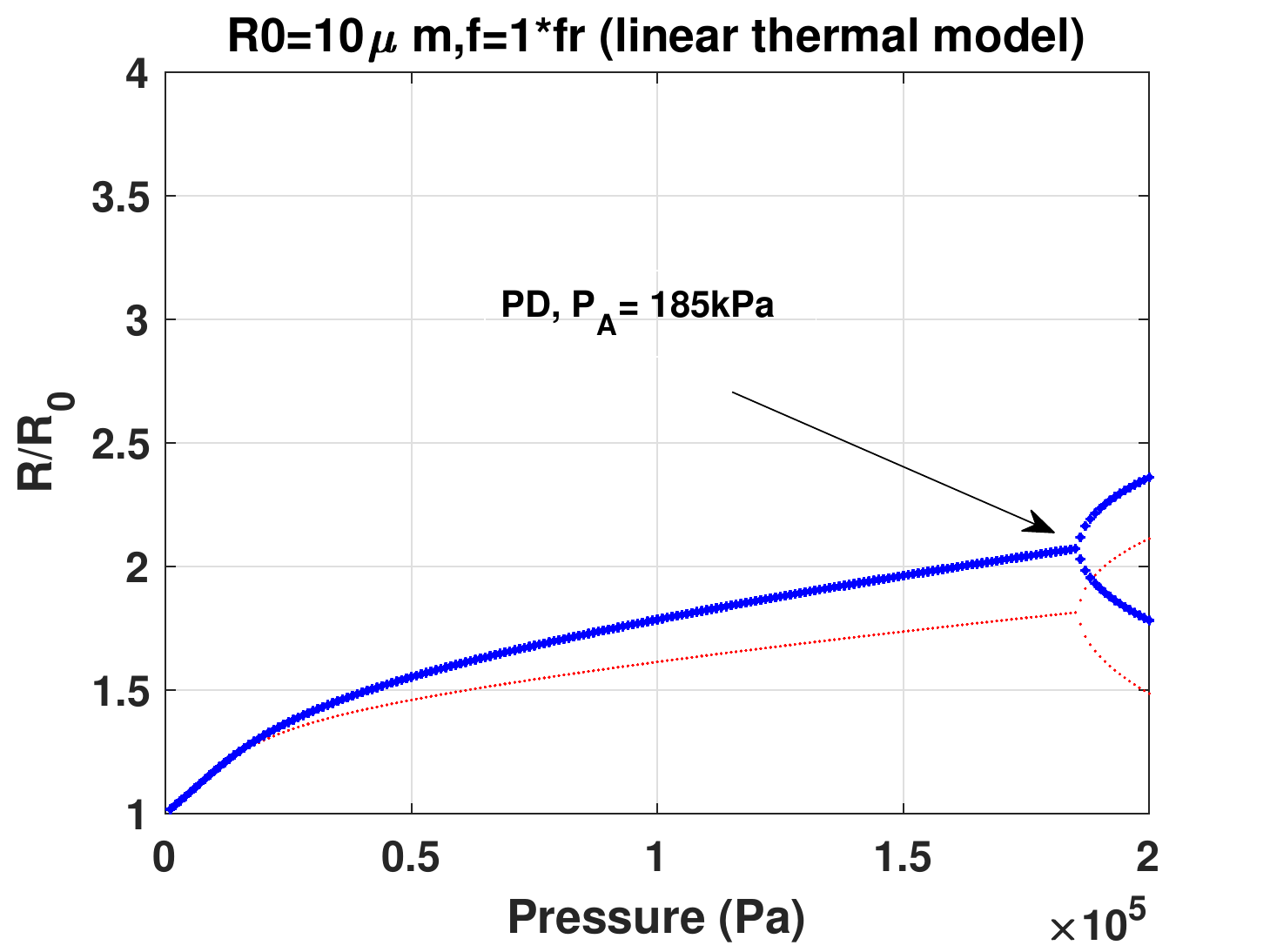}}	\scalebox{0.3}{\includegraphics{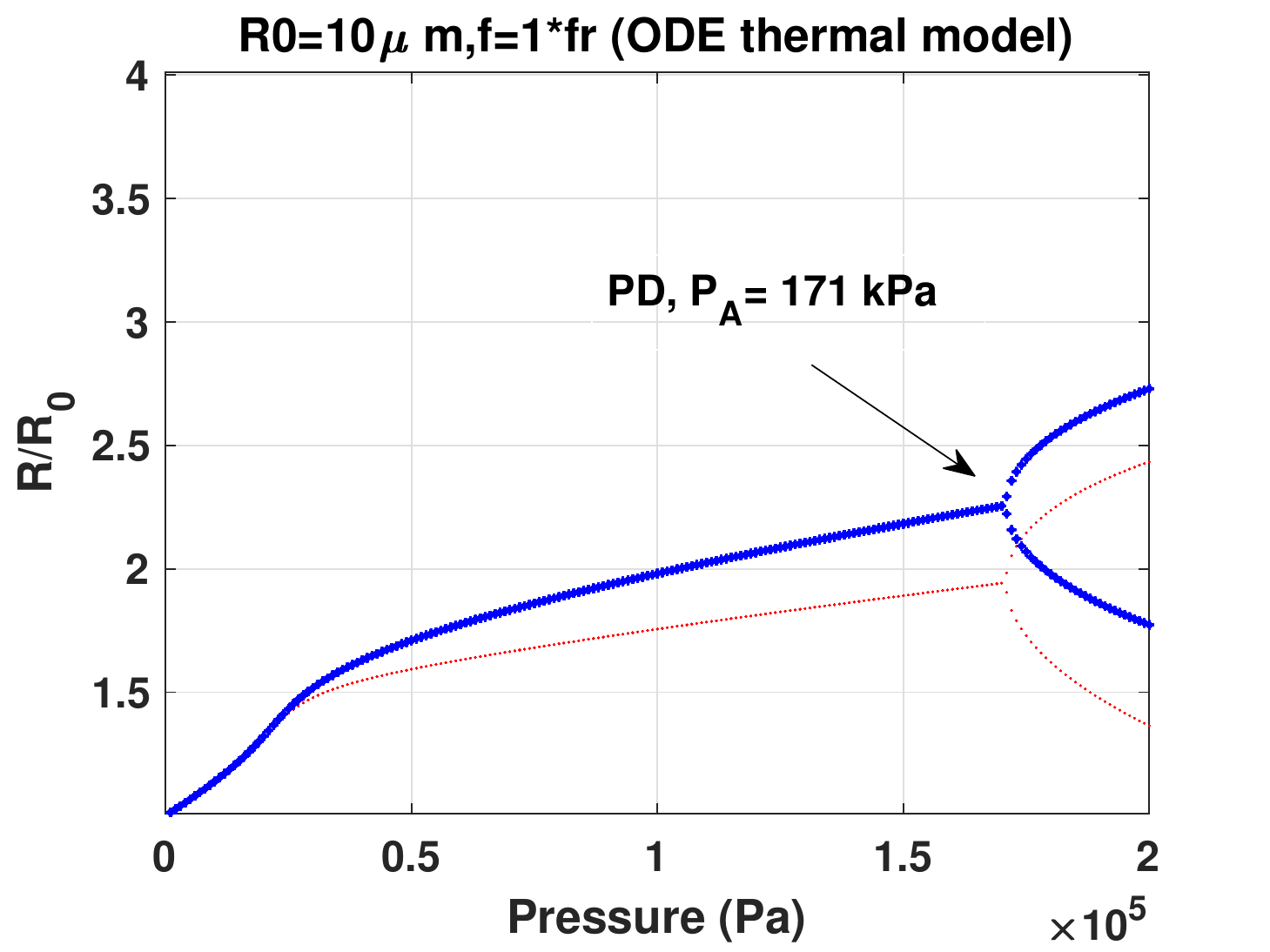}}\\
		(d) \hspace{3.5cm} (e) \hspace{3.5cm} (f) \\
		\scalebox{0.3}{\includegraphics{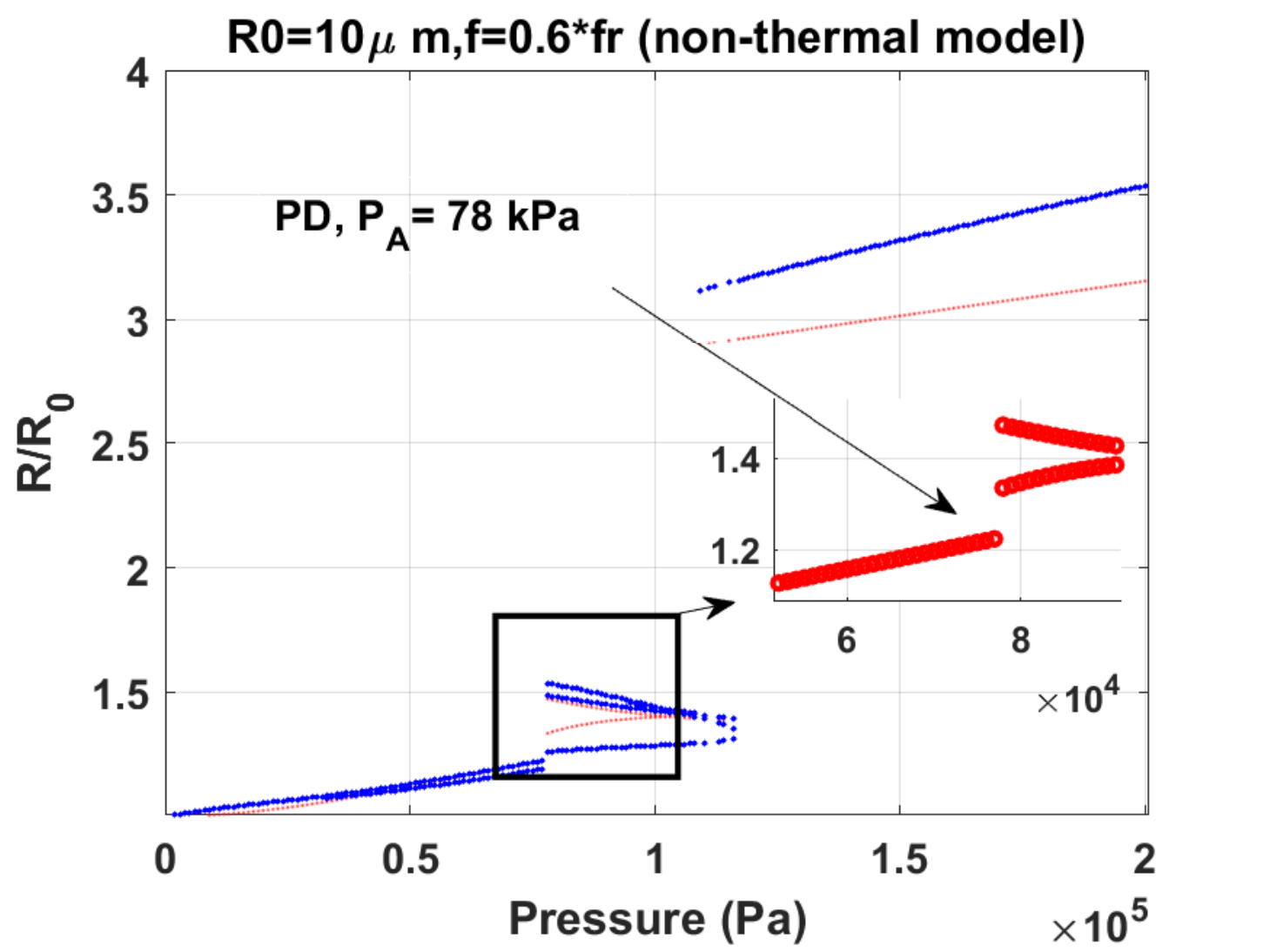}}\scalebox{0.3}{\includegraphics{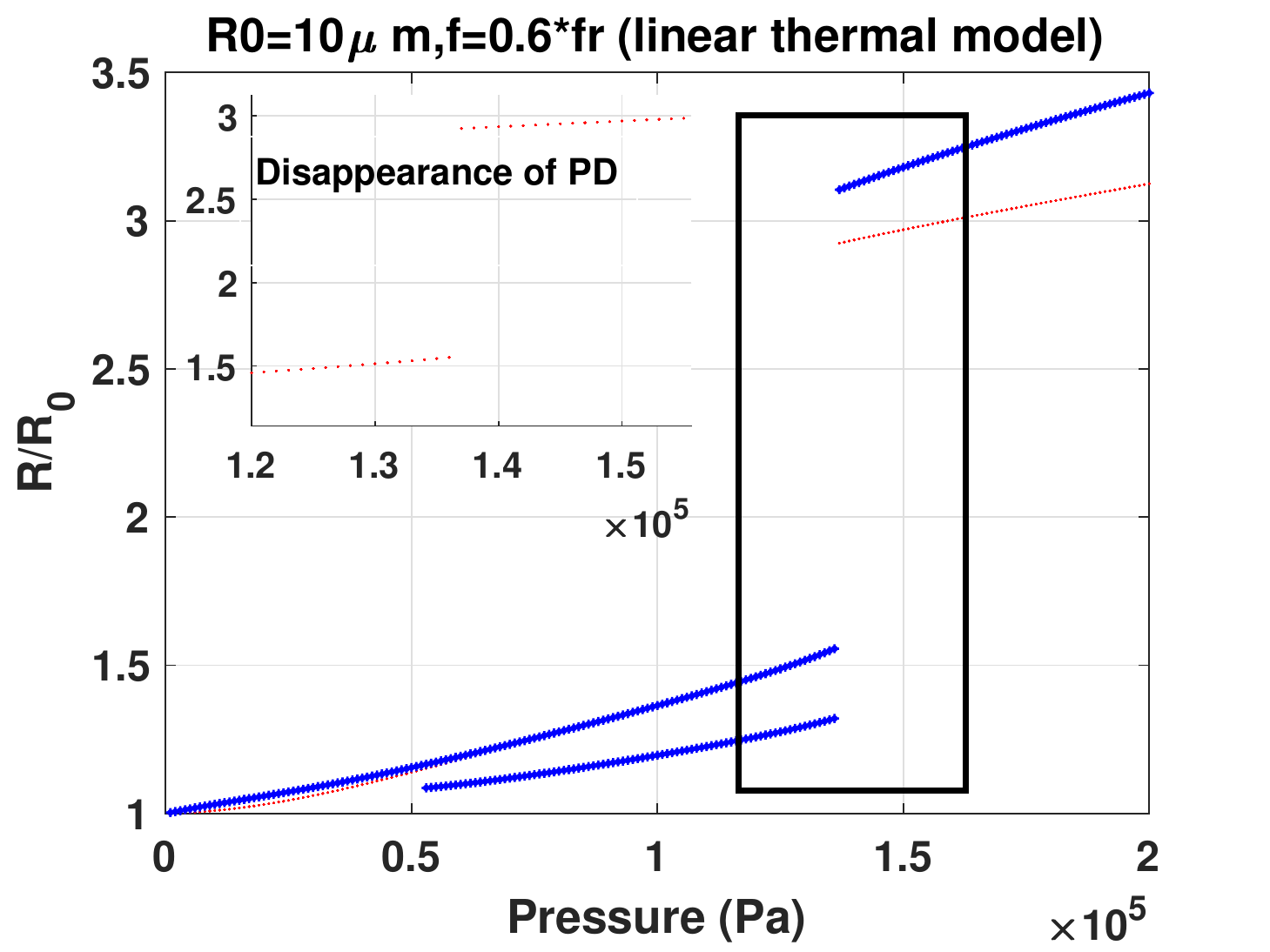}}	\scalebox{0.3}{\includegraphics{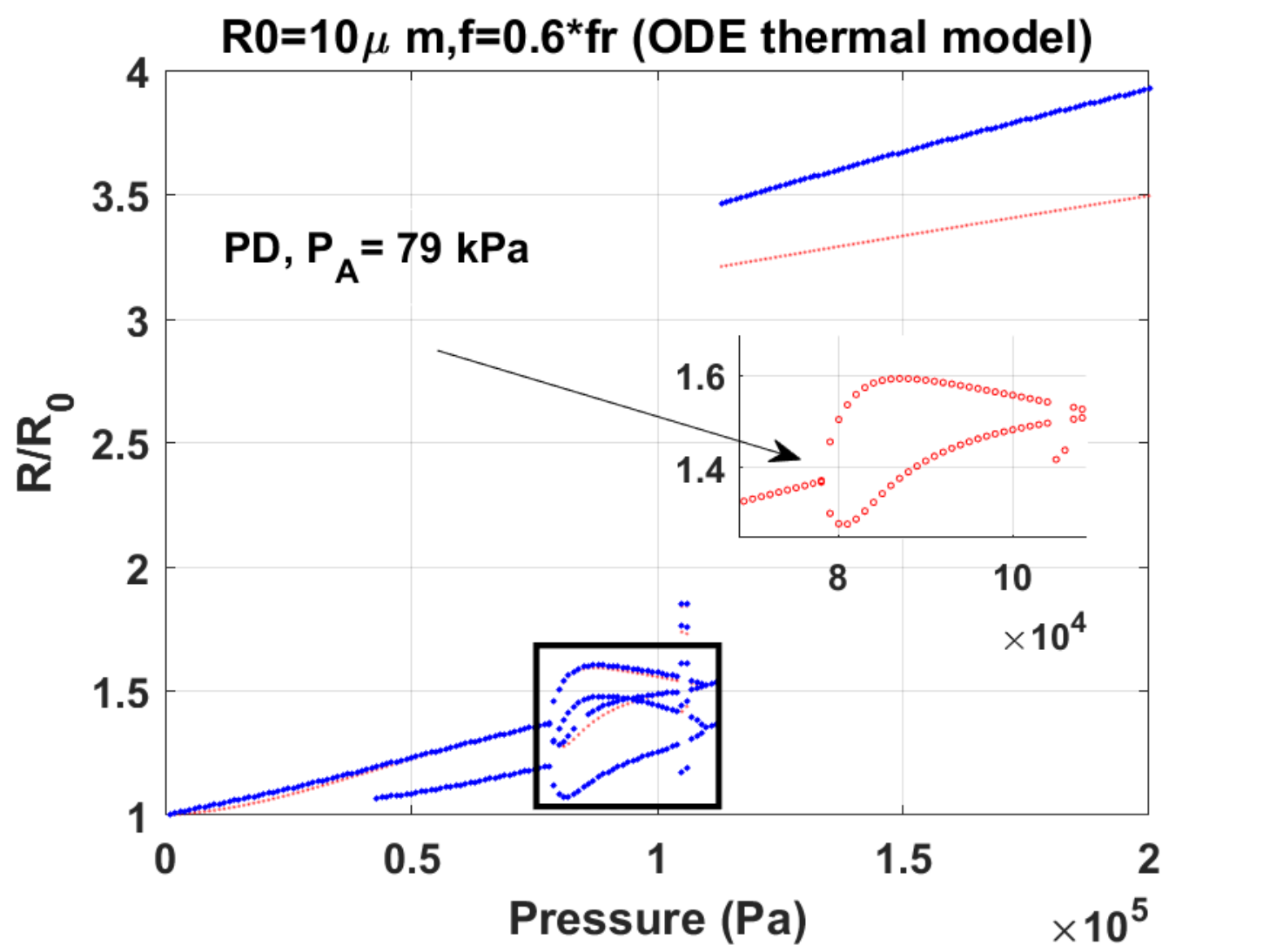}}\\
		(g) \hspace{3.5cm} (h) \hspace{3.5cm} (i) \\
		\caption{Bifurcation structure of the $R/R_0$ of an air bubble with $R_0=10 \mu m$ as a function of pressure (The red curves are generated through the Poincaré method and the blue curves are generated via the method of maxima).Top row: $f=2f_r$, middle row: $f=f_r$ and bottom row: $f=0.6f_r$.}
	\end{center}
\end{figure*}

\begin{figure*}
	\begin{center}
		\scalebox{0.3}{\includegraphics{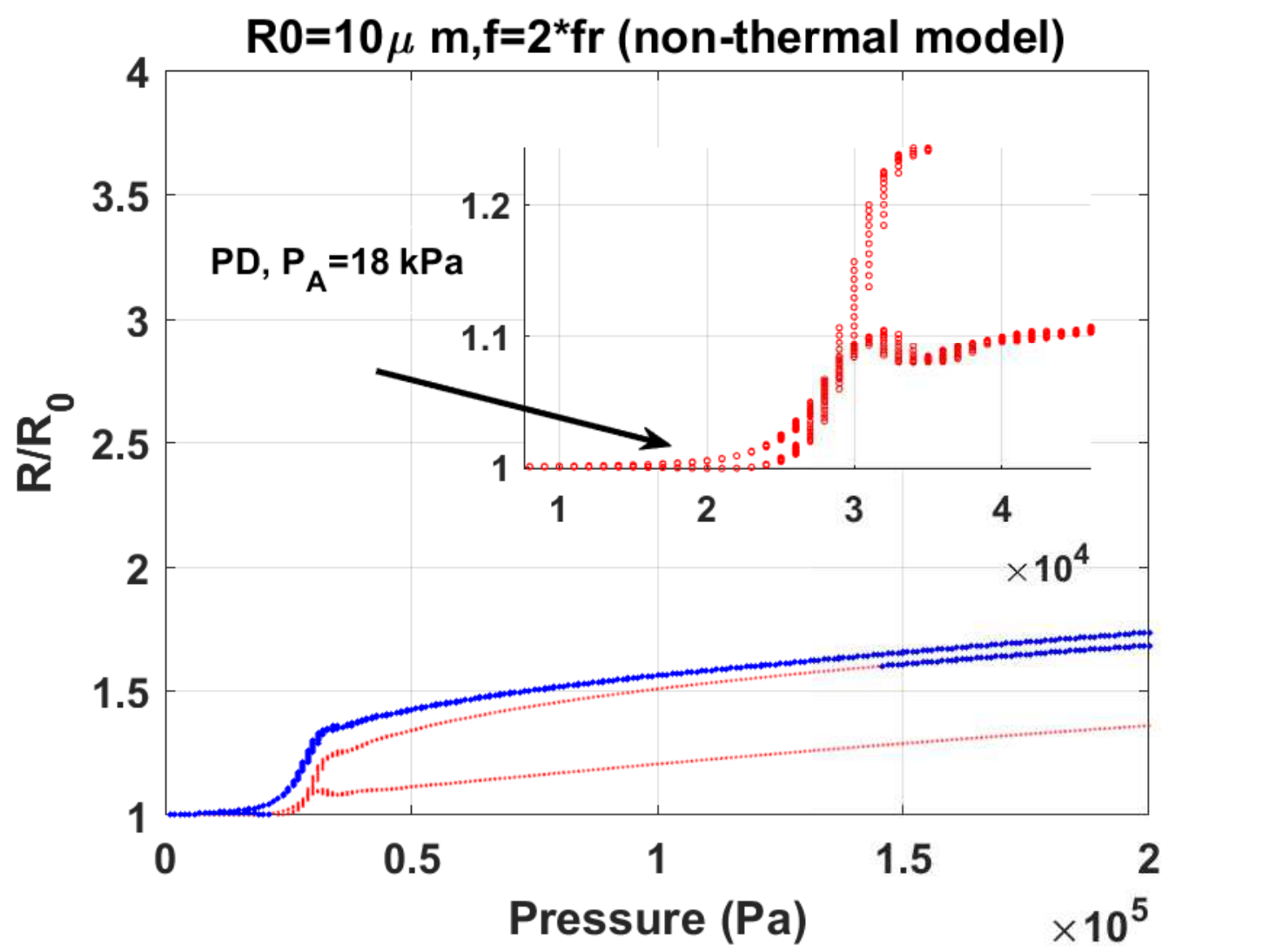}}\scalebox{0.3}{\includegraphics{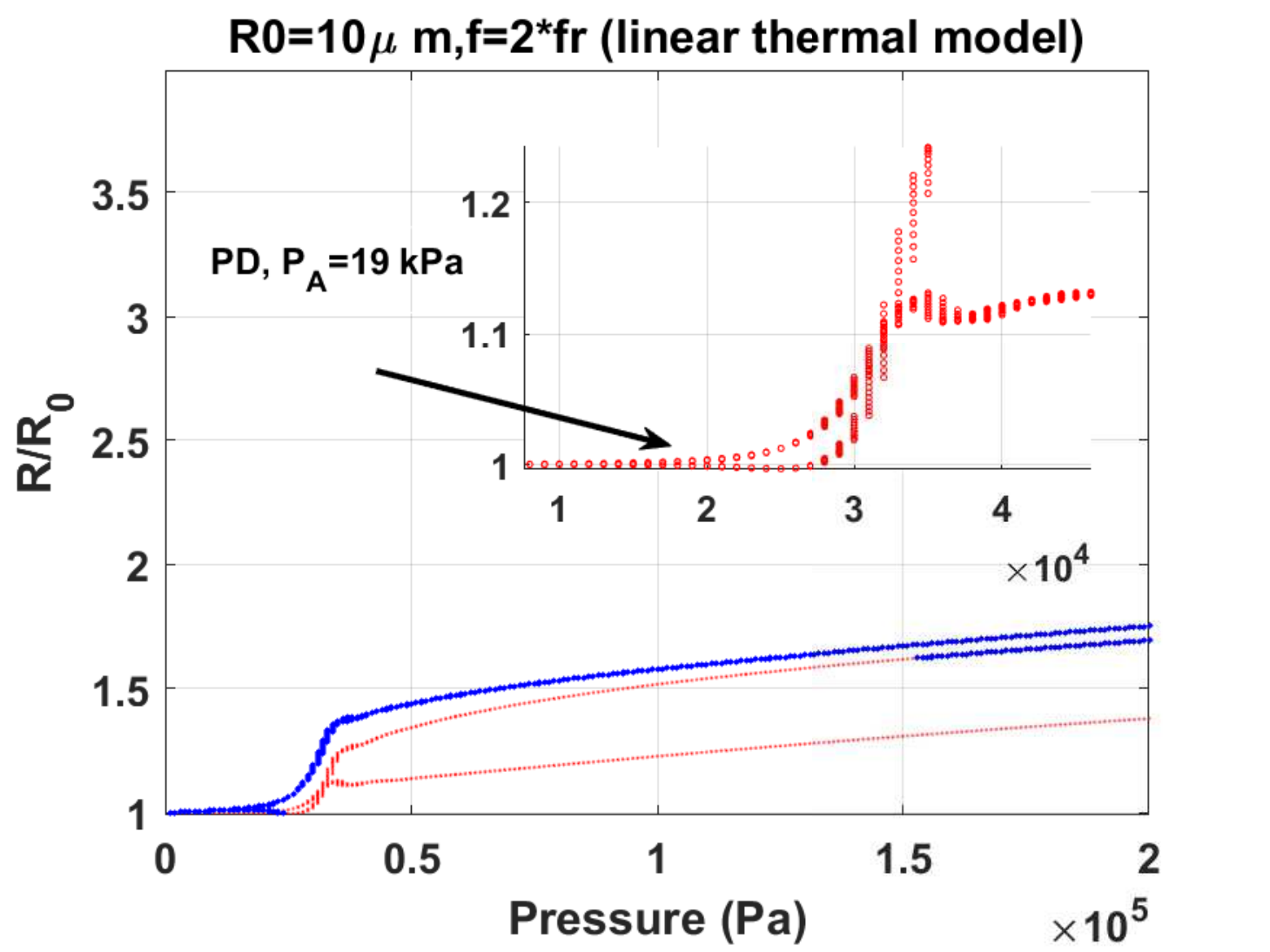}}	\scalebox{0.3}{\includegraphics{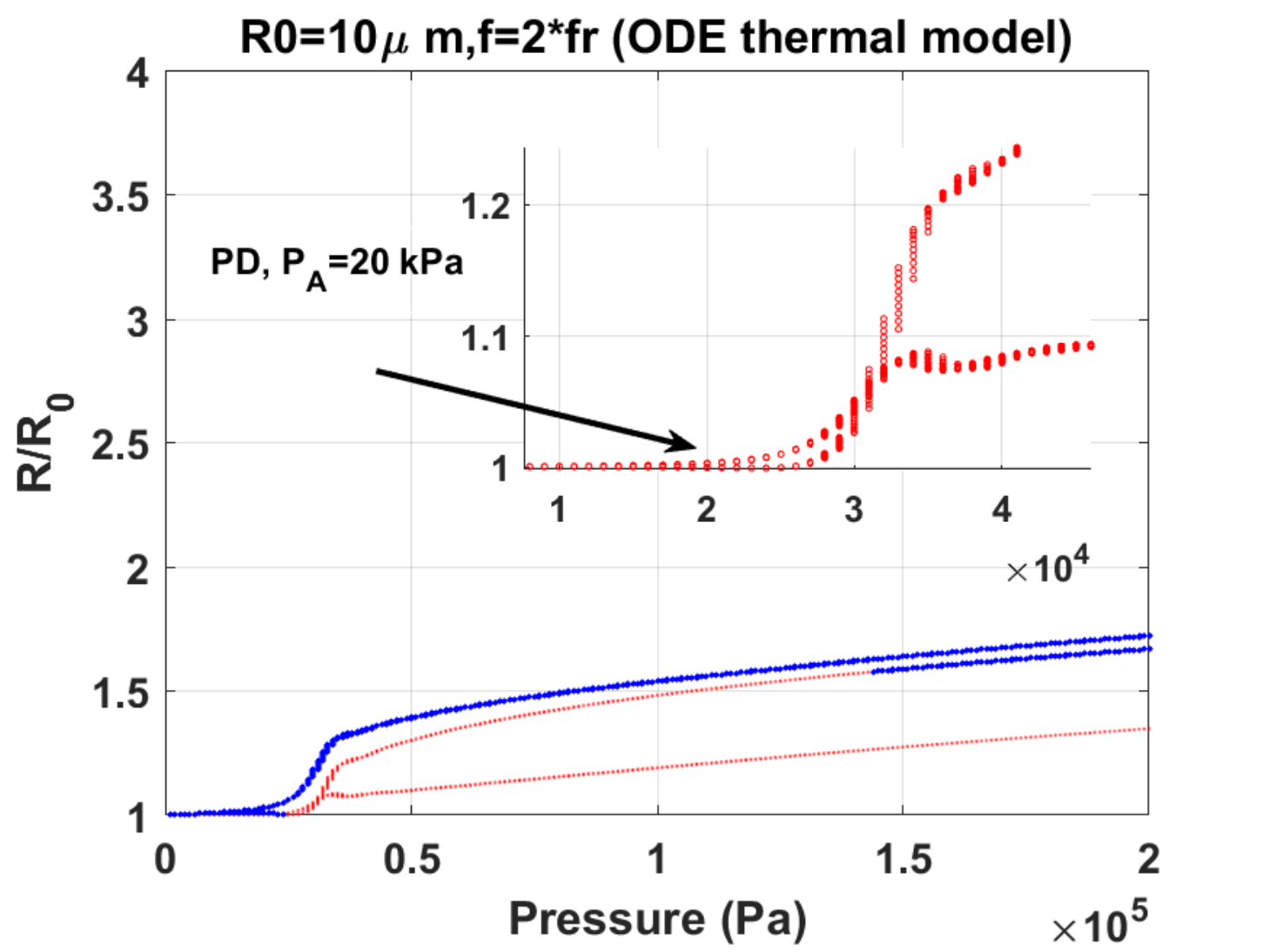}}\\
		(a) \hspace{3.5cm} (b) \hspace{3.5cm} (c) \\
		\scalebox{0.3}{\includegraphics{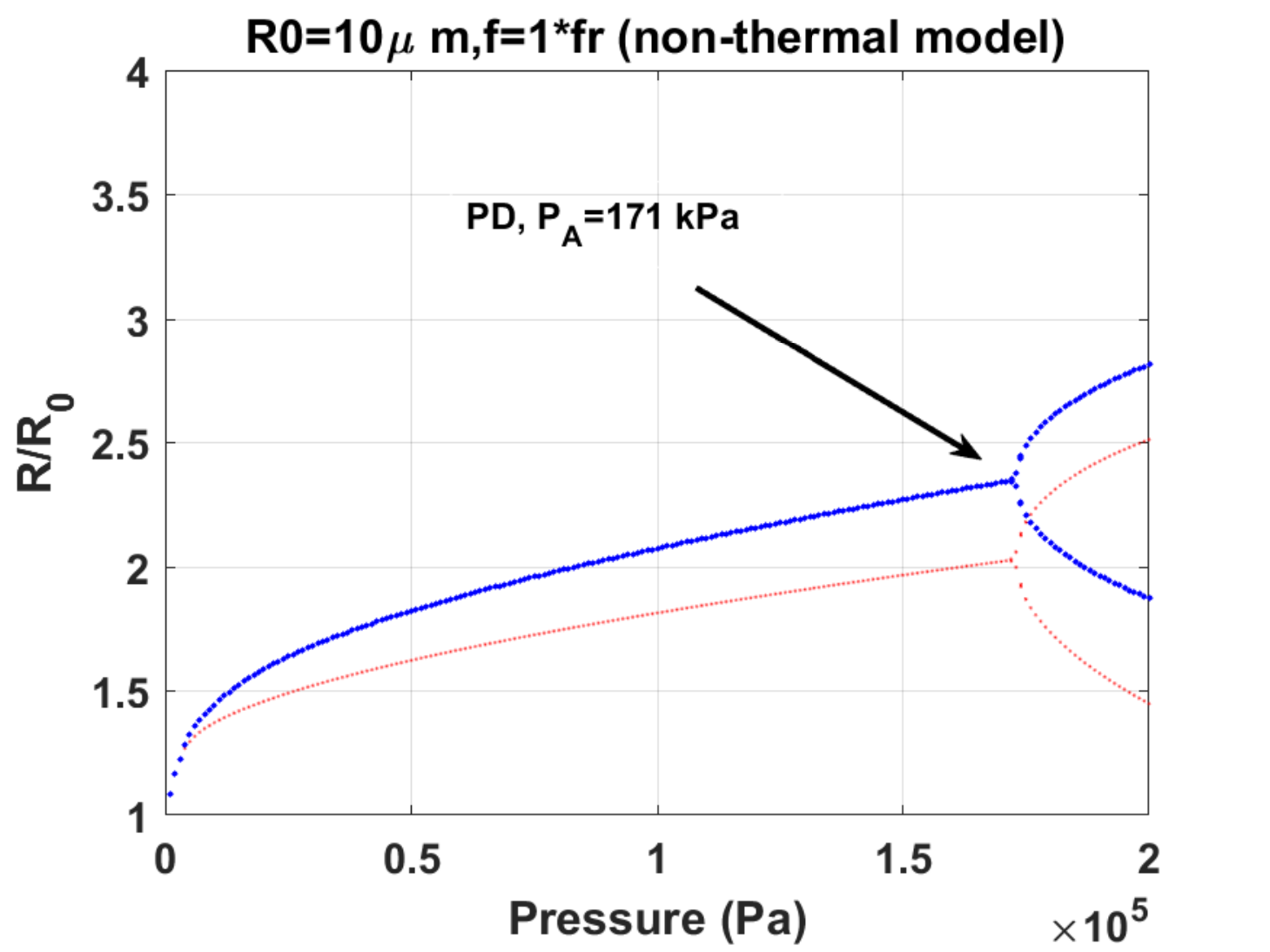}}\scalebox{0.3}{\includegraphics{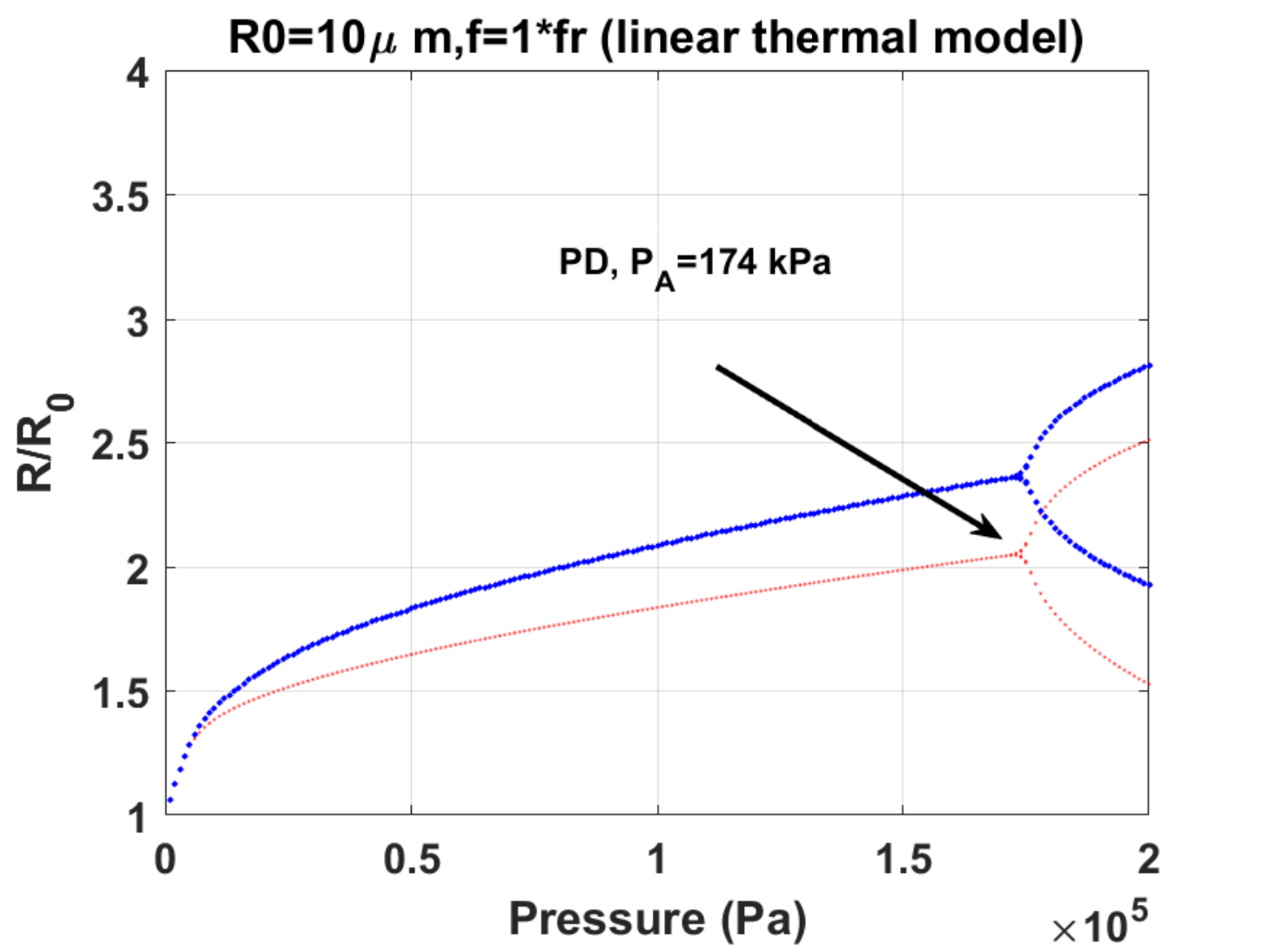}}	\scalebox{0.3}{\includegraphics{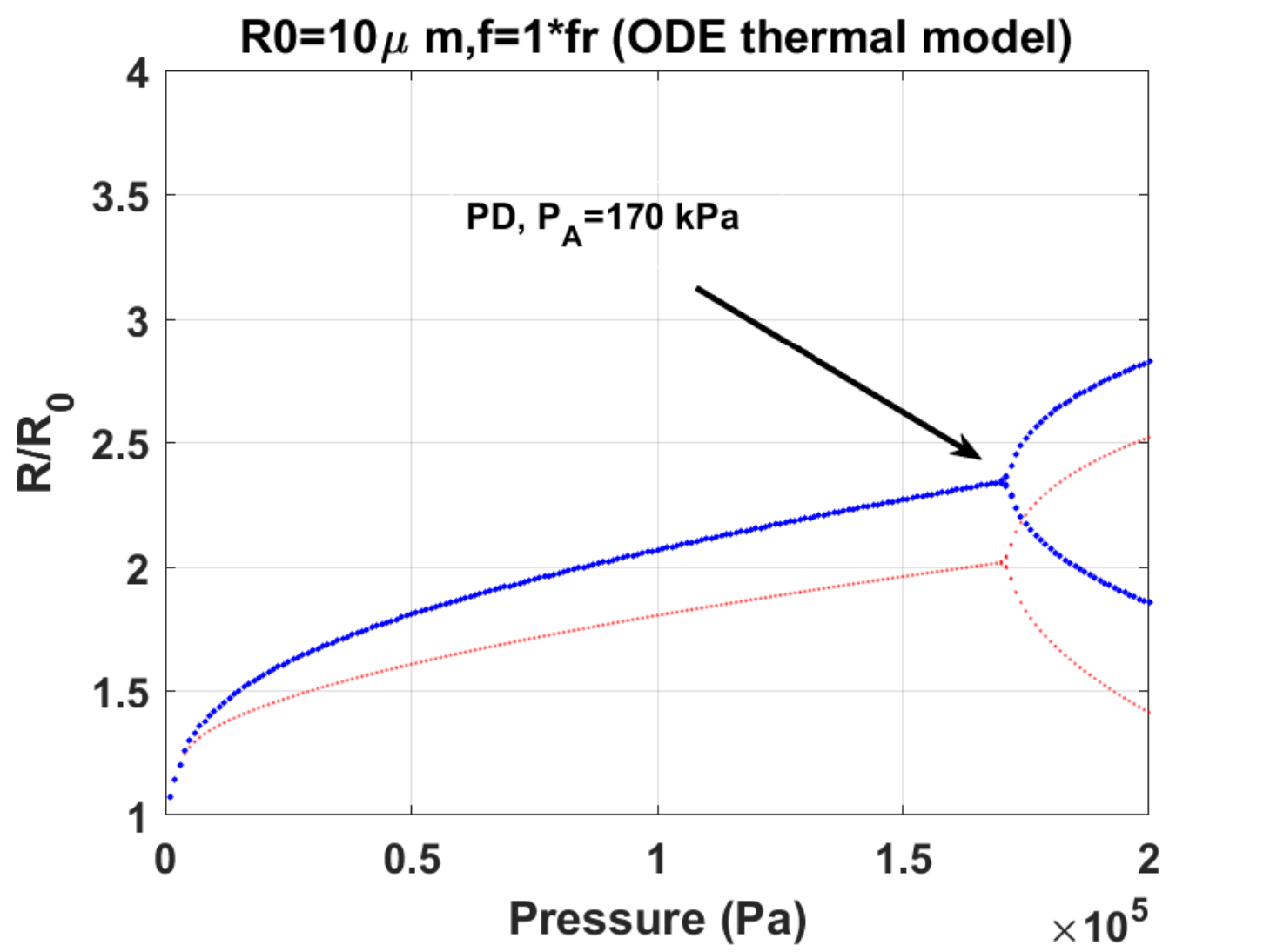}}\\
		(d) \hspace{3.5cm} (e) \hspace{3.5cm} (f) \\
		\scalebox{0.3}{\includegraphics{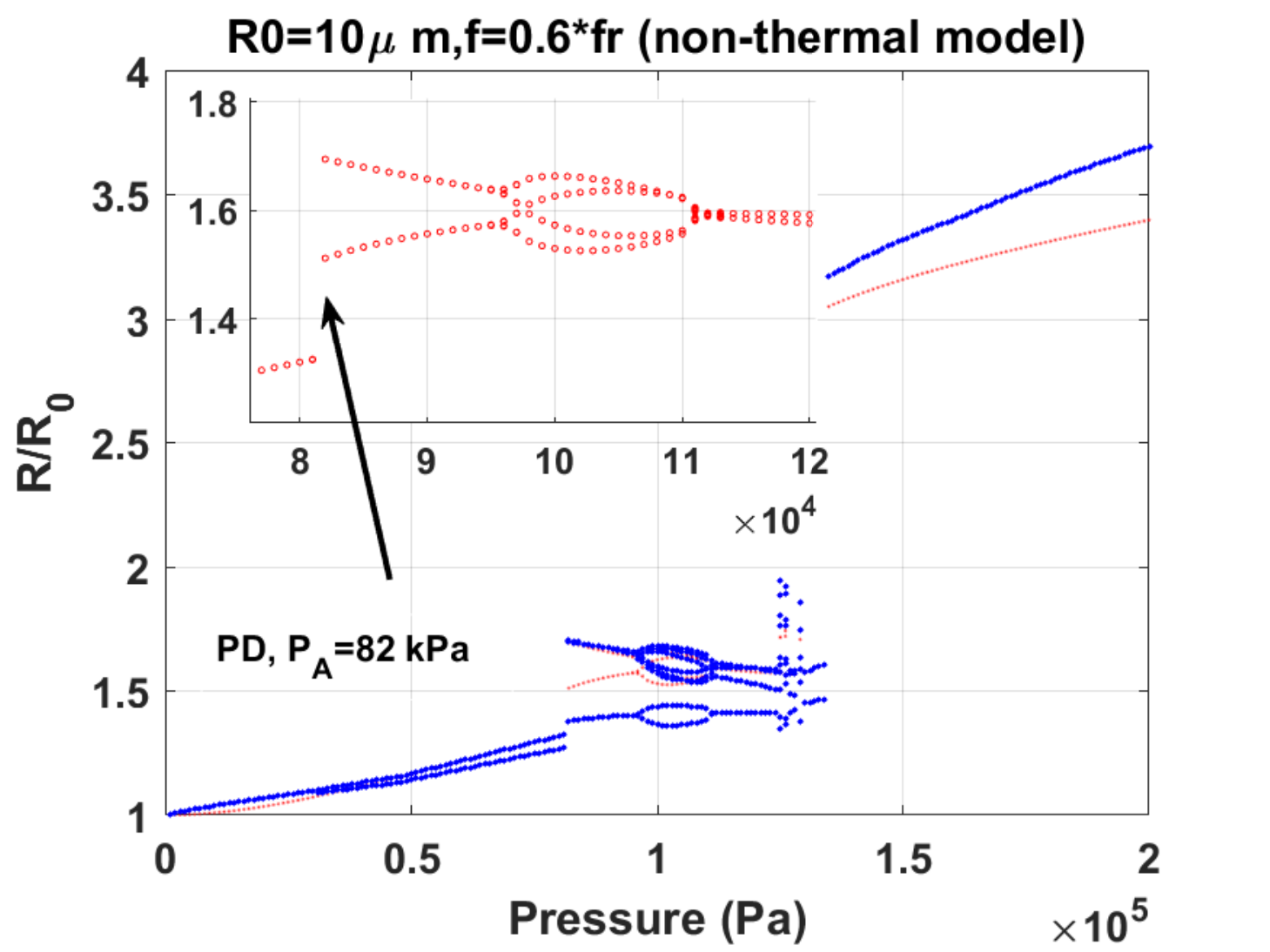}}\scalebox{0.3}{\includegraphics{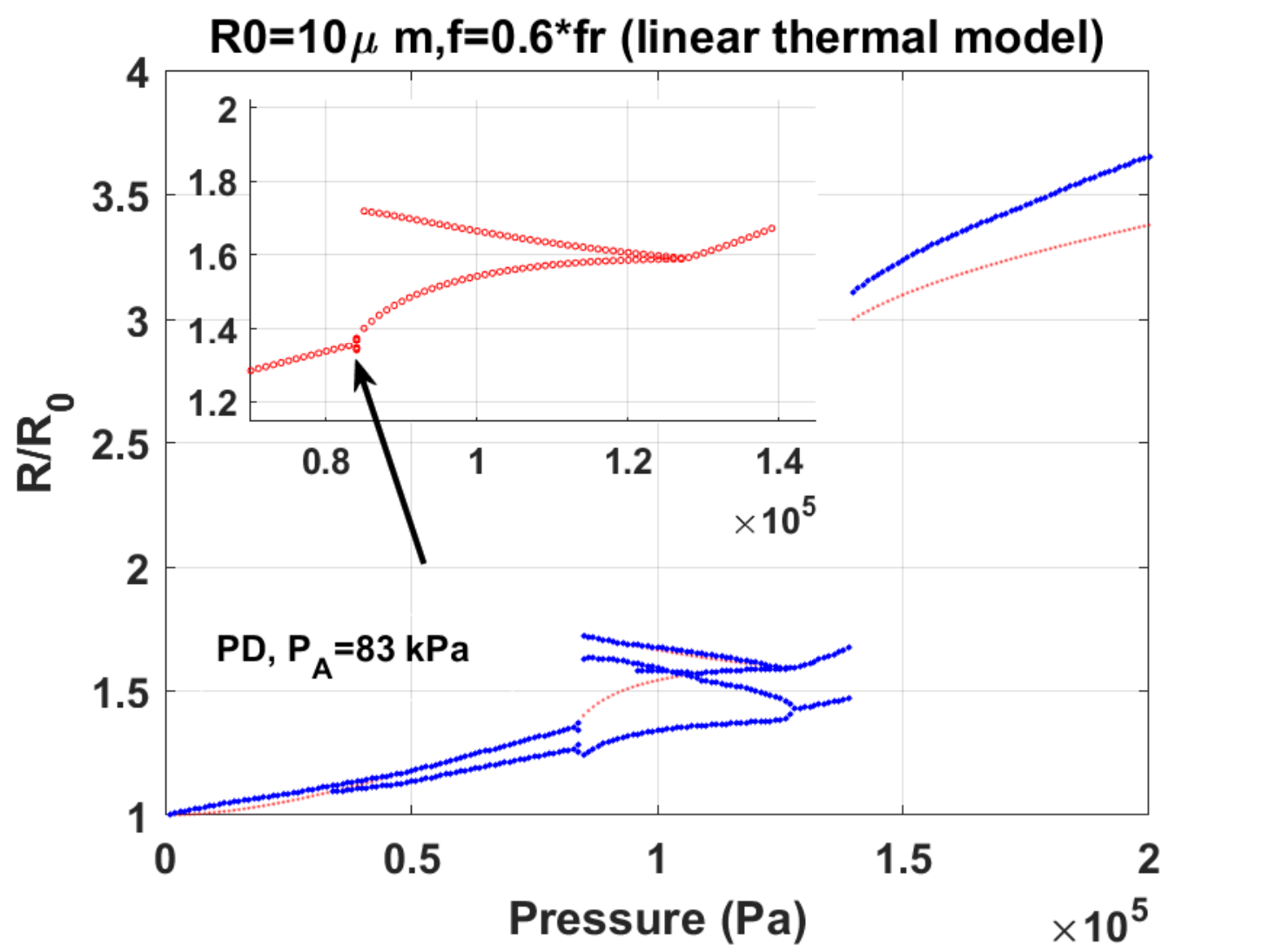}}	\scalebox{0.3}{\includegraphics{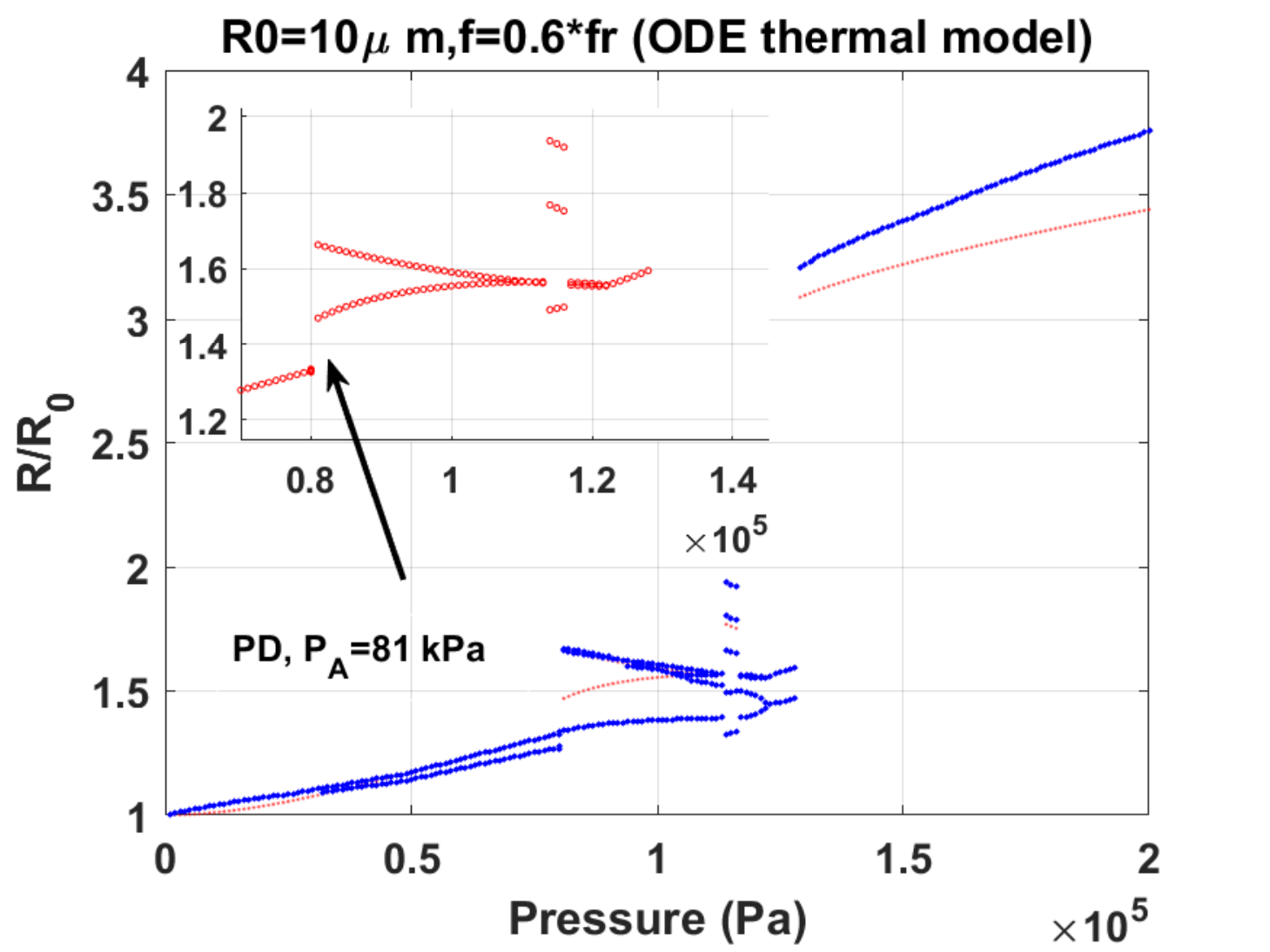}}\\
		(g) \hspace{3.5cm} (h) \hspace{3.5cm} (i) \\
		\caption{Bifurcation structure of the $R/R_0$ of a C3F8 bubble with $R_0=10 \mu m$ as a function of pressure (The red curves are generated through the Poincaré method and the blue curves are generated via the method of maxima). Top row: $f=2f_r$, middle row: $f=f_r$ and bottom row: $f=0.6f_r$.}
	\end{center}
\end{figure*}
\section{Results and Discussion}
\subsection{Some samples of the bifurcation structure}
Figure 1 shows the bifurcation structure of an air bubble with $R_0=10 \mu m$ as a function of pressure. The frequency of sonication are in multiples of resonance frequency ($f_r$). For each model the resonance frequency is calculated numerically by solving the corresponding equations at $P_A=1kPa$ and finding the frequency of the maximum radial expansion. The first column in Fig. 1 is calculated by solving Eq. 1 and neglecting the thermal effects (non-thermal model). The results in the second column take into account the linear thermal dissipation by coupling Eq. 1 with Eqs. 4-7 (linear thermal model) and the third column considers the full thermal effects through coupling Eq.1 with Eqs. 2 and 3 (full thermal model). The red graph is generated through the Poincare method (2.3.1) and the blue graph is generated via the method of maxima (2.3.2).\\ 
Fig. 1 shows that in the case of the non-thermal model (NTM) and linear thermal model (LTM), the minimum pressure threshold for PD occurs when $f=2f_r$ (Figs. 1a and 1b). However, when full thermal effects are considered the minimum pressure threshold is when $f=0.6f_r$ (Fig. 1i).\\  When $f=2f_r$ (Figs. 1a-c) the bubble undergoes period doubling (PD) at $P_A=17 kPa$, $48 kPa$ and $96 kPa$ predicted by NTM, LTM and FTM respectively. Period doubling is highlighted in a subset in each graph where the red graph bifurcates in to two branches. The FTM model predicts the highest pressure threshold for the generation of PD and SH emissions. We have recently shown \cite{56,57} that linear models for dissipation effects in bubble oscillations may loose accuracy at acoustic excitation pressures \textit{as low as 20 kPa}.  Predictions of the the LTM and the FTM deviate as pressure increases and the FTM should be used to model bubble oscillations \cite{57,58}. This is because the LTM is derived assuming very small (1-2 percent expansion ratio), and symmetric bubble oscillation amplitude. However, as pressure increases, the radial oscillations grow above the size limit where linear assumptions are valid and the bubble expansion and compression phases are no longer symmetric. Higher internal temperatures can be created during significant compression or collapse (larger $R_{max}/R_{min}$) and the average surface area available for thermal conduction becomes smaller or larger depending on the regime of oscillations \cite{56,57}. This causes the predictions of the LTM and FTM to diverge. When $f=2f_r$ the LTM model under-predicts the PD threshold. This is because it treats the behavior of the thermal dissipation similar to viscous dissipation through adding a thermal viscosity ($\mu_{th}$) term (Eq. 6). Thus, thermal dissipation behaves similarly to liquid viscous dissipation ($\sim \frac{4\mu_{th}\dot{R}}{R}$) which is proportional to the wall velocity amplitude. However, when $f=2f_r$, for bubbles with $R_0>0.7 \mu m$, PD occurs at very gentle wall velocities ($\approx<$ 5 m/s \cite{6}) and thus the thermal dissipation as predicted by the LTM model will be small. When PD occurs at $f=2f_r$, the bubble average surface area available for thermal conduction increases, and subsequently thermal dissipation (Td) increases. Thus, as shown in \cite{57}, compared to radiation damping and liquid viscosity damping, Td exhibits the largest increase when PD occurs at $f=2f_r$ (e.g. Fig. B.1.g in \cite{56}). When PD occurs at $f=2f_r$, the wall velocity is small, and the bubble rebounds slowly after collapse (\cite{6}). After the occurence of the PD, the radial oscillations have one maximum (the bubble only collapses once every two acoustic cycles). Even when the second maxima reappears at higher pressures, its amplitude is very close to that at the first maxima. Due to the large surface area in each period and slow rebound after every collapse, the average surface area for thermal conduction is larger, and thus the Td as predicted by the FTM model is higher than the LTM. Thus, in case of the bubble with $R_0=10 \mu m$ (Fig. 1c) due to the increased Td, FTM predicts a higher threshold for PD.\\ When $f=f_r$, PD occurs at 174 kPa, 185 kPa and 171 kPa as predicted by NTM, LTM and FTM models respectively. In this case PD occurs at a higher pressure in the case of the LTM model. This is because when $f=f_r$ due to the very large amplitude bubble velocities at PD (for $R_0>0.7 \mu m$, $|\dot{R}|>35 m/s$), thermal dissipation is over estimated by the LTM model as Td $\sim \frac{4\mu_{th}\dot{R}}{R}$. However, in reality the average surface area for temperature escape becomes limited. When $f=f_r$ and at PD, the oscillations have two distinct maxima. The bubble collapses strongly (leading to increased temperature), and it rebounds back to a radius that is smaller than the previous maxima. Thus, the bubble surface are becomes smaller. Moreover, the bubble rebounds quicker, thus the time duration available for the temperature escape will be limited. These lead to reduction of the average surface area for thermal conduction and thus Td decreases.  FTM model incorporate these effects. Therefore, it predicts a lower pressure threshold for PD. The FTM model also predicts a slightly lower pressure threshold (171 kPa in Fig. 1f) for the generation of PD tcompared to the NTM model (174 kPa in Fig. 1d). Our detailed analysis of the pressure dependent mechanisms in \cite{58} reveals that when $f=f_r$, the radiation damping (Rd), thermal damping (Td) and damping due to liquid viscosity ($Ld$) increase with pressure increase with the order of strength of $Td>Ld>Rd$. However, Rd grows faster than the other dissipation factors with increasing pressure. When PD occurs, due to the decrease in average surface area for thermal conduction, Td decreases and Rd becomes the major dissipation mechanism ($Rd>Td>Ld$). In case of the NTM model and due to the absence of Td, the bubble attains higher wall velocities and acceleration and due to the dominant role of Rd at PD ($Rd \sim (R\ddot{R}+2\dot{R}^2$))  the total dissipation increases slightly which lead to a 3 kPa higher pressure for the generation of PD.\\
When $f=0.6f_r$, the P1 oscillation amplitude increases with increasing pressure and a second maximum (two blue lines) appears above a pressure threshold of $\approx 40 kPa$. This leads to the enhancement of the 2nd harmonic in the pressure scattered by the bubble \cite{46}.  Above a second pressure threshold, PD occurs (red line bifurcates to two branches) at 78 kPa (Fig. 1g) and 79 kPa (Fig. 1i) respectively for the NTM and FTM model. The LTM model (Fig. 1h) \textit{does not predict any PD} due to overestimation of the Td. Overestimation of the Td in case of the LTM is due to the very large wall velocities that the bubble attain in the regime of 5/2 UH emissions \cite{58}. The P2 oscillation has 4 maxima (Fig. 1d and 1f) indicating that the 5/2 UH component of the scattered pressure is stronger than other SH and UH components \cite{46}.\\  Fig. 2 shows the bifurcations structure of a C3F8 bubble with $R_0=10 \mu m$ as a function of pressure. In case of C3F8 gas core, due to the significantly weaker effects of Td \cite{56,59}, the 3 models predict the same behavior for the bubble. Thus in agreement with previous analytical and numerical studies that negelcet or simplify Td,  the lowest pressure threshold for the generation of PD is at $f=2f_r$.\\  
The results indicate that the minimum pressure for the generation of PD strongly depends on the Td. For gases like air that have high Td, the frequency of minimum pressure threshold for PD can shift to frequencies below resonance.
\begin{figure*}
	\begin{center}
		\scalebox{0.3}{\includegraphics{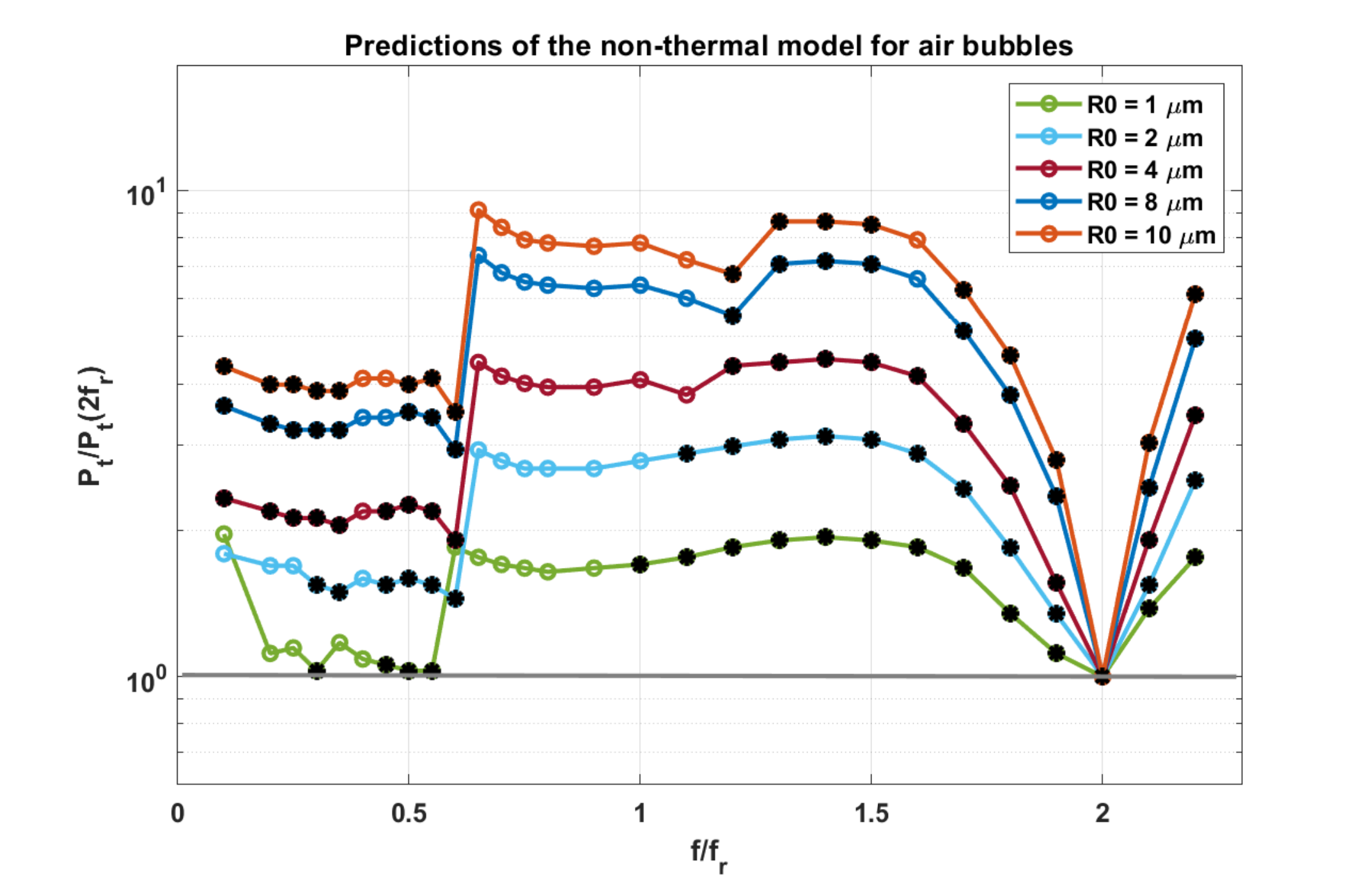}} \scalebox{0.3}{\includegraphics{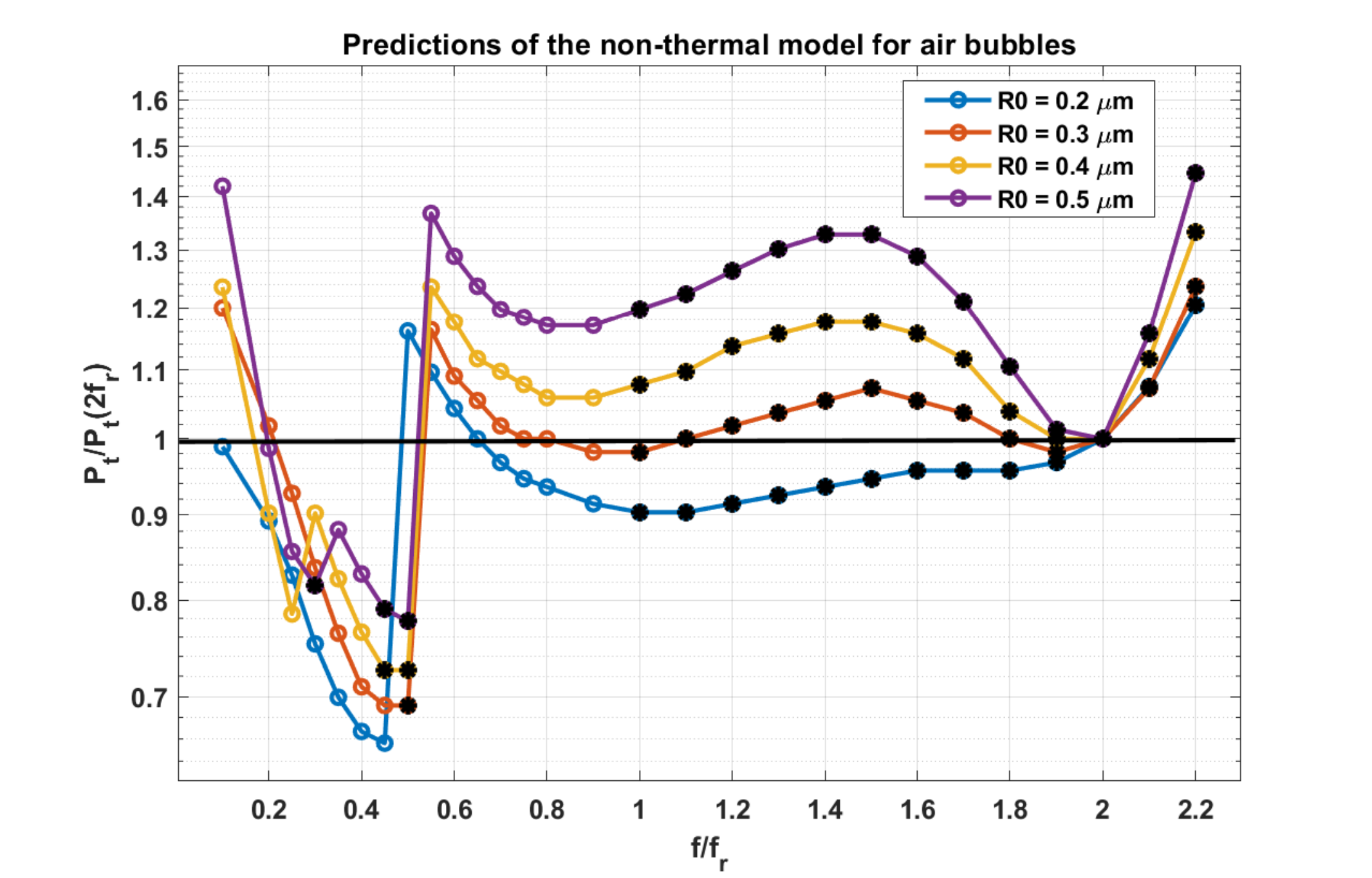}}\\
		\hspace{0.5cm} (a) \hspace{6cm} (b)\\
		\scalebox{0.3}{\includegraphics{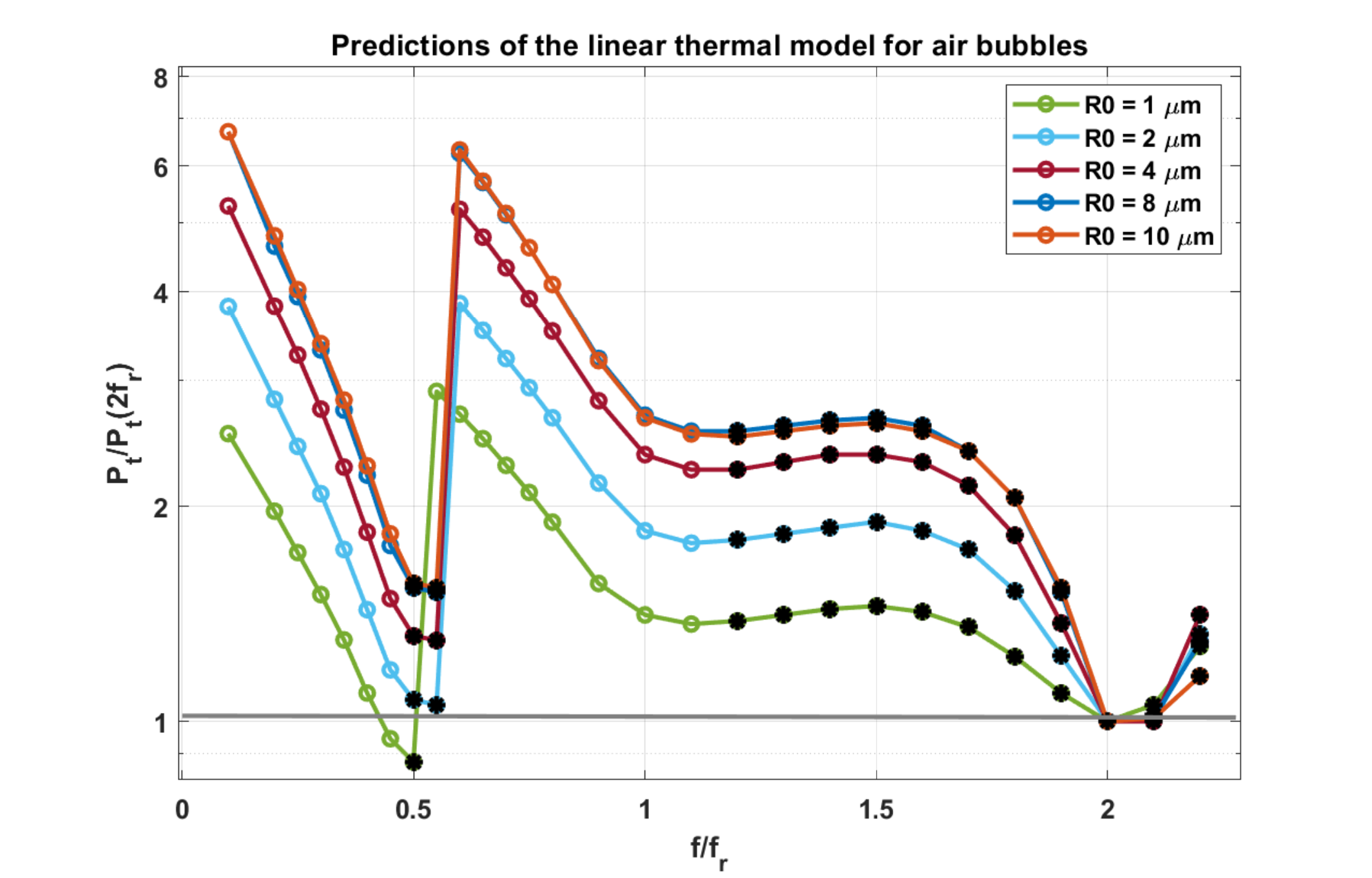}} \scalebox{0.3}{\includegraphics{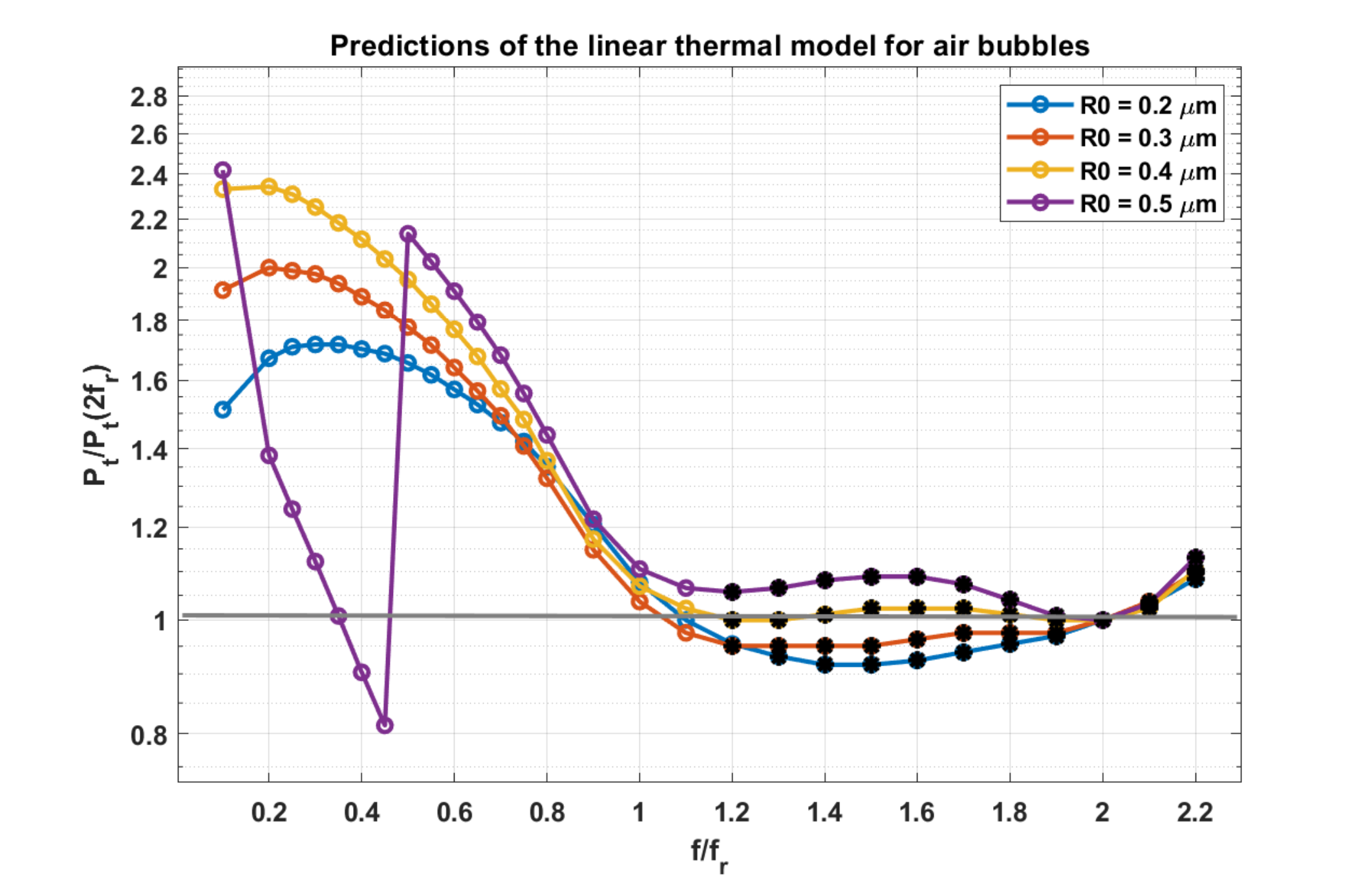}}\\
		\hspace{0.5cm} (c) \hspace{6cm} (d)\\
		\scalebox{0.3}{\includegraphics{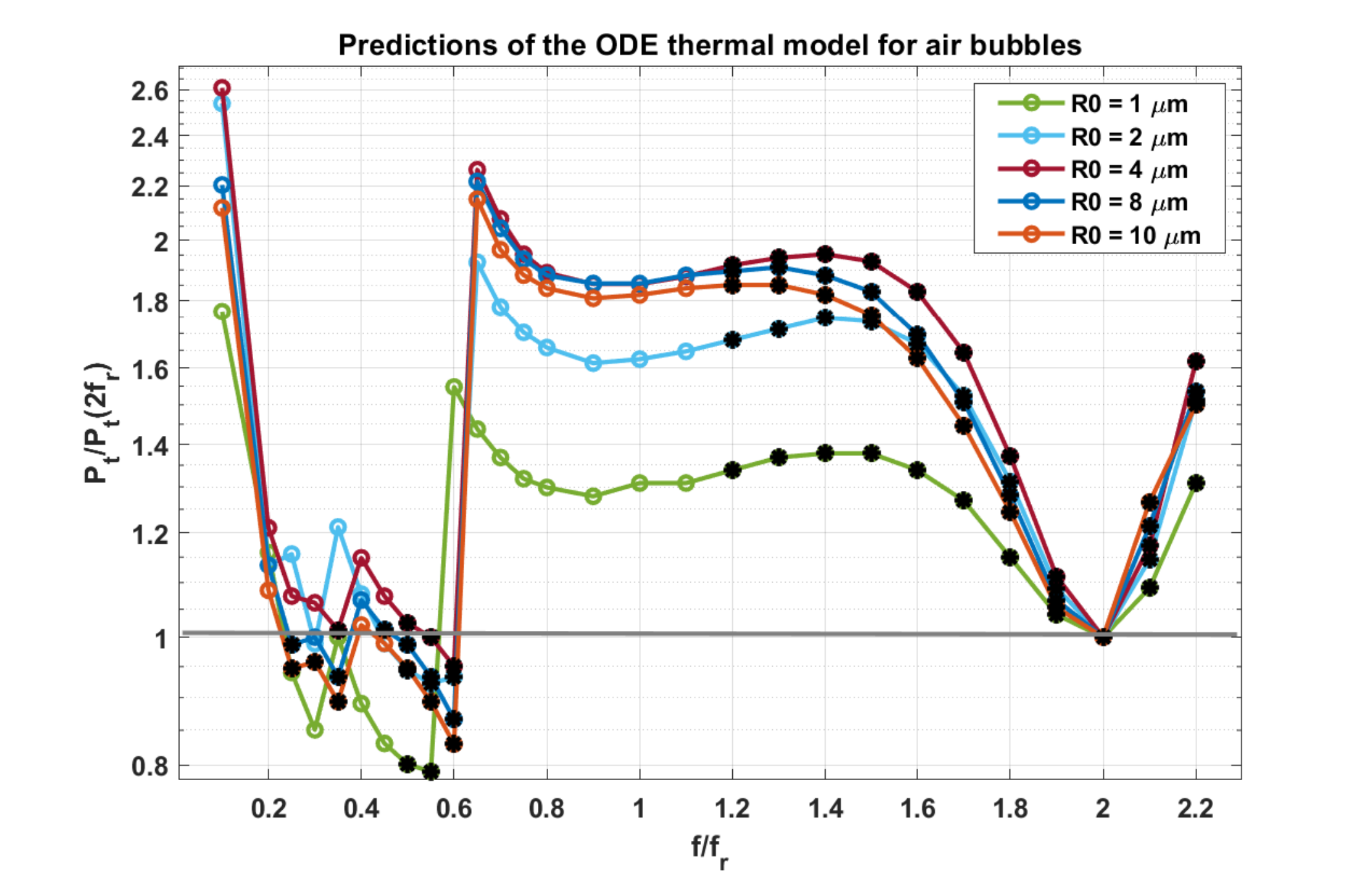}} \scalebox{0.3}{\includegraphics{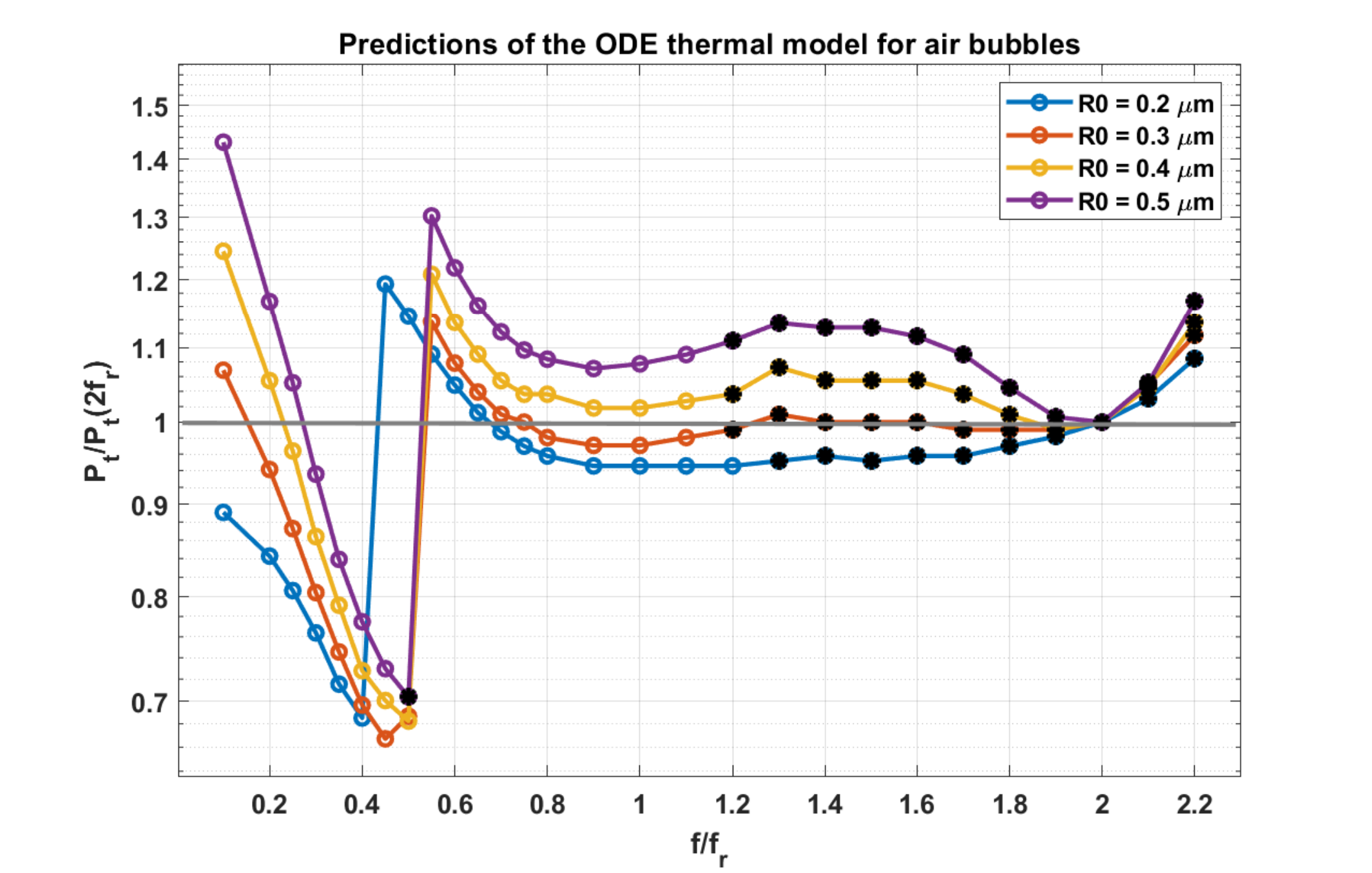}}\\
		\hspace{0.5cm} (e) \hspace{6cm} (f)\\
		\caption{Normalized pressure thresholds of the P2 oscillations of an air bubble as a function of frequency. The left column is for bubbles with initial radii between $1 \mu m\leq R_0 \leq 10 \mu m$ and the right column is for bubbles with initial radii between $ 0.2 \mu m\leq R_0 \leq 0.5 \mu m$: (a) $\&$ (b)- Non-thermal model, (c) $\&$ (d)-linear thermal (a) $\&$ (f)- ODE thermal model.  Solid black circles mark the stable oscillation regimes.}
	\end{center}
\end{figure*}
\subsection{Pressure threshold of P2 oscillations as a function of size and frequency}
In this section the pressure threshold for the generation of the PD is extracted from the bifurcation diagrams and plotted as a function of frequency. The  PD pressure threshold is \textit{normalized to the pressure threshold at $f=2f_r$} for simplicity of comparison. Moreover, the regions of non-destructive oscillations are identified.  Stable domains are identified as oscillations that satisfy the relationship of $R_{max}/R_0\leq 2$  \cite{60,61,62}. As it is reviewed in \cite{6,62}, it appears that the minimum reported threshold for bubble destruction. In order to ensure that potential bubble destruction does not take place and to assure adverse bio-effects are avoided, using the minimum reported bubble destruction threshold is reasonable. The regions for nondestructive bubble oscillations are marked with solid black circles in Figs. 3 and 4.\\
The results of this section are presented in two categories for each of the NTM, LTM and FTM models for better readability and distinction of physical effects. The first category is for bubbles with initial radii of $1 \mu m\leq R_0\leq 10\mu m$ for which thermal effects are stronger. The second category is for bubbles with $0.2 \mu m\leq R_0\leq 0.5\mu m$ for which the viscous effects are stronger. Fig. 3a shows that for air bubbles with $1 \mu m\leq R_0\leq 10\mu m$ and in the absence of Td, the minimum pressure threshold for PD occurs at $f=2f_r$. This is in agreement with the results of analytical studies \cite{34,35,36,37,38} and numerical simulations \cite{39,40} that neglect the thermal dissipation. For the studied frequency range, the NTM model predicts stable SH emissions for $0.3f_r<f<0.7f_r$ and $f\geq1.2f_r$. For bubbles with $0.2 \mu m\leq R_0\leq 0.5\mu m$ however, the minimum pressure threshold occurs for $0.5fr\leq f\leq 0.6f_r$. Previous numerical studies \cite{39,40} showed that the minimum pressure threshold for SH generation shifts from twice the resonance frequency toward the resonance frequency for bubbles with $R_0\leq 0.3 \mu m$. These studies conclude that, increased damping dampens the bubble response more at twice the resonance than at the resonance which leads to the shift of the frequency of minimum pressure threshold. The results presented in Figs. 3b, 3d and 3f are in agreement with these studies and predict a minimum near the resonance frequency for bubbles with $R_0\leq 0.3 \mu m$ in the frequency range of $0.7f_r<f<2.2f_r$. However, the previous studies did not cover frequencies below $0.7f_r$ and thus were not able to reveal the absolute minimum that exists in $0.5fr\leq f\leq 0.6f_r$. The occurrence of PD in this frequency range is due to 5/2 UH resonance (e.g. Fig. 1g (P2 with 4 maxima)). The  NTM model predicts that the SH emissions are stable for $f\geq1.2f_r$ and $0.2f_r<f<0.7f_r$ (solid black circles).\\
When linear thermal dissipation is considered, bubbles with $1 \mu m\leq R_0\leq 10\mu m$ exhibit a minimum at $2f_r$, except the bubble with $R_0=1\mu m$. The minimum threshold for SH emissions for the bubble with $R_0=1 \mu m$ is at $f=0.5f_r$. Since the smaller bubbles are more affected by viscous damping, the addition of the thermal viscosity ($\mu_{th}$) to the liquid viscosity in the LTM model significantly increases the total damping. Thus, the bubble behavior with $R_0= 1\mu m$ becomes  similar to the behavior of the smaller bubbles with very high viscous damping (Fig. 3b). The shift in the minimum frequency in this case, thus, can be described with a similar explanation given in \cite{38,39}. For bubbles with $0.2\mu m\leq R_0\leq0.5 \mu m$ the minimum pressure threshold occurs for frequencies near resonance, except for the bubble with $R_0=0.5 \mu m$. The addition of the $\mu_{th}$ to the liquid viscosity significantly increase the total dissipation in smaller bubbles (as Td becomes inversly proportional to bubble radius) and makes the bubble oscillator an over-damped oscillator. In this case, UH oscillations are suppressed and the PD requires much higher pressures for $f<f_r$. Moreover the P2 oscillations are not in non-destructive regime for $f\leq 1.1f_r$.  For bubbles with $0.2 \mu m\leq R_0\leq 0.5\mu m$, the LTM predicts non-destructive SH emissions only for $1.2f_r \leq f_r\leq 2.2f_r$.\\
When thermal effects are considered with their full non-linearity and for bubbles with $1 \mu m\leq R_0\leq 10\mu m$ (Fig. 3e), minimum pressure threshold for SH emission is when $0.5f_r<f\leq0.6f_r$. SH emissions are stable for $f\geq 1.2f_r$ and $0.5f_r\leq f\leq 0.6f_r$. The bigger bubble ($R_0=10 \mu m$) may sustain non-destructive SH emissions for $0.25f_r\leq f\leq 0.35f_r$ as well. For bubbles with $0.2\mu m\leq R_0\leq0.5\mu m$ (Fig. 3f) the minimum pressure threshold for SH emissions occurs at $0.4f_r\leq f\leq0.5f_r$, however, only the bubble with $R_0=0.5f_r$ may exhibit stable SH emissions.\\ 
In case of FTM, all the bubbles in this size range exhibit stable SH emissions when $1.2f_r\leq f\leq 2.2f_r$.\\
 \begin{figure*}
	\begin{center}
		\scalebox{0.3}{\includegraphics{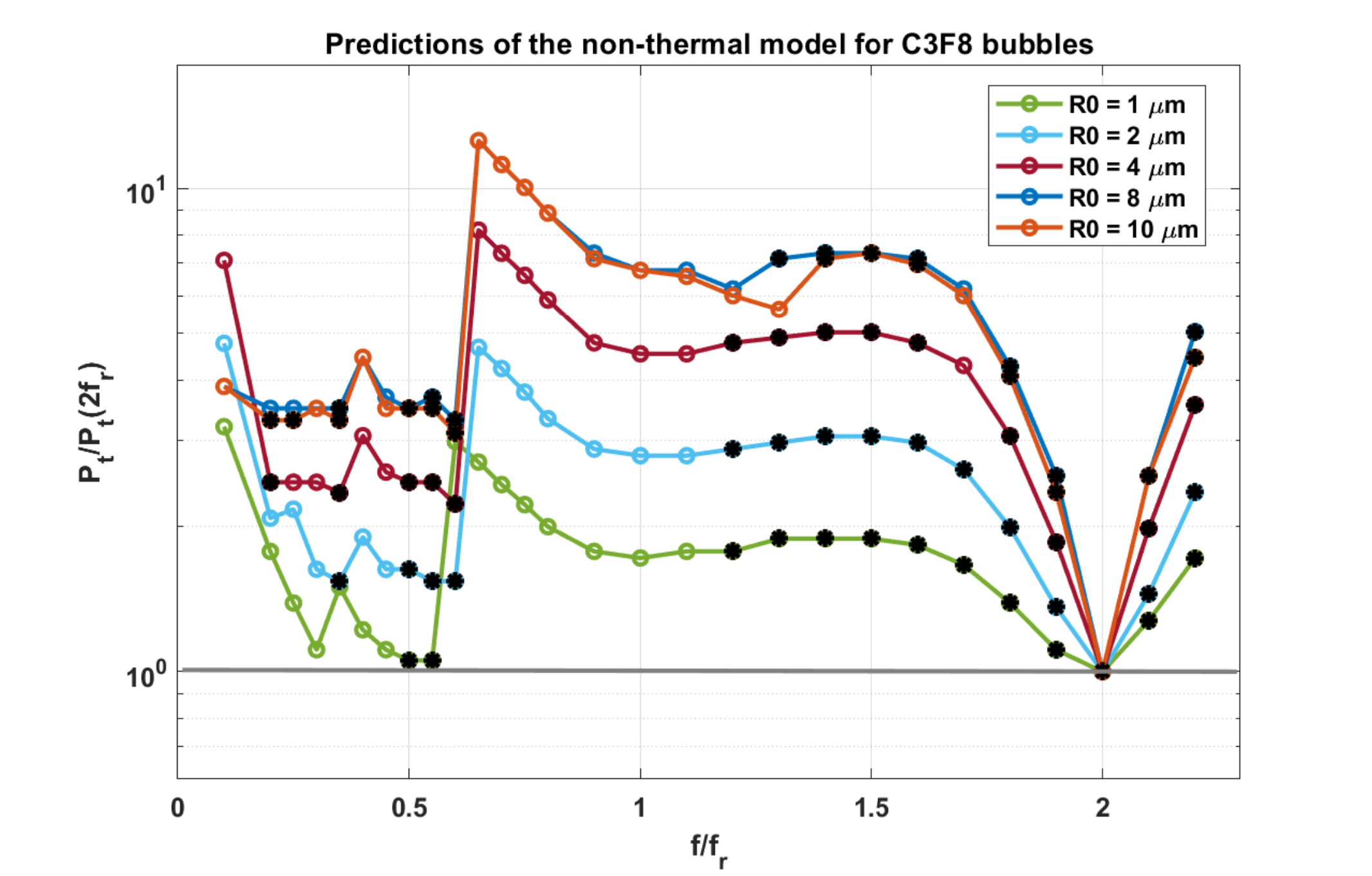}} \scalebox{0.3}{\includegraphics{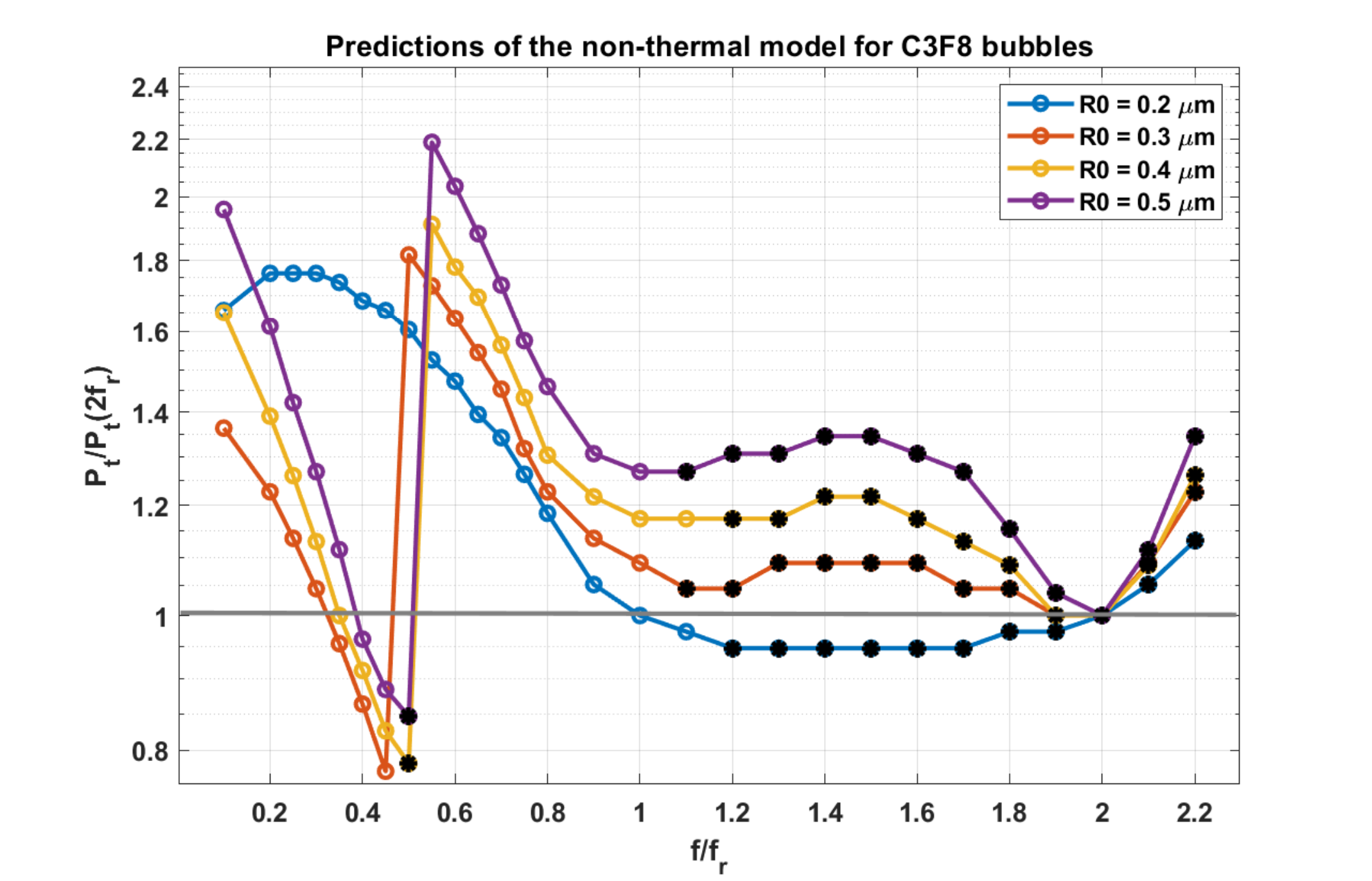}}\\
		\hspace{0.5cm} (a) \hspace{6cm} (b)\\
		\scalebox{0.3}{\includegraphics{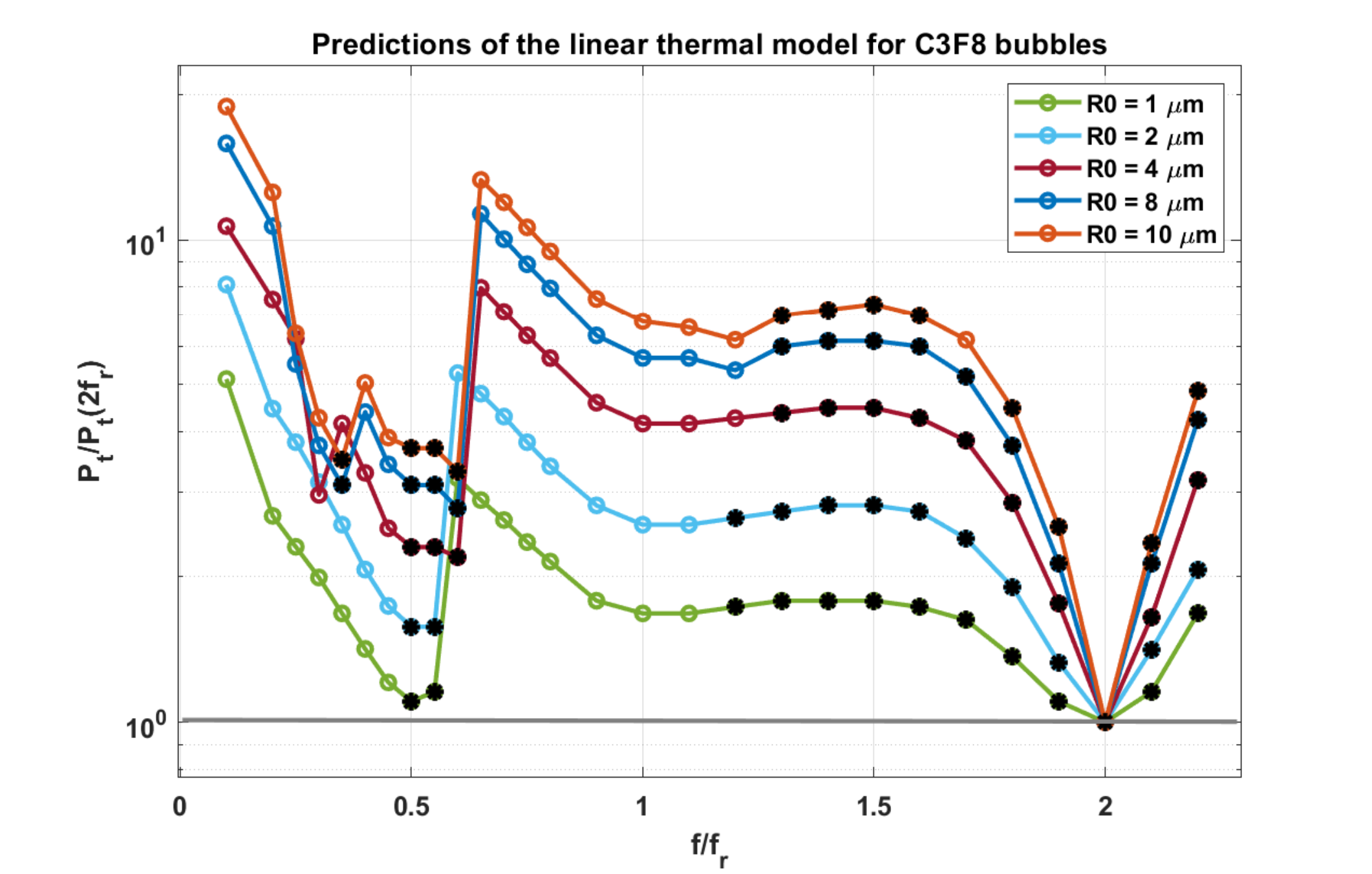}} \scalebox{0.3}{\includegraphics{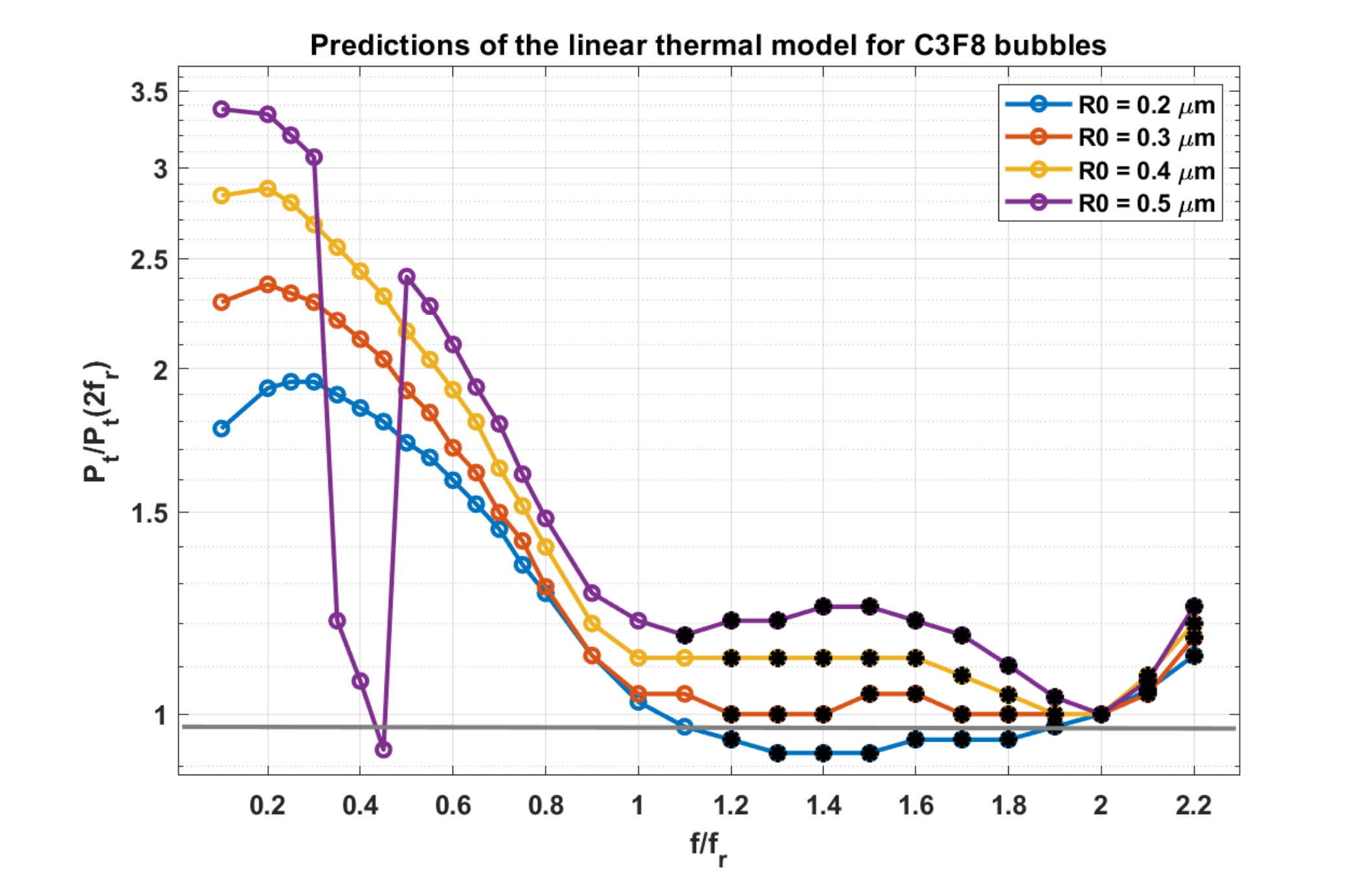}}\\
		\hspace{0.5cm} (c) \hspace{6cm} (d)\\
		\scalebox{0.3}{\includegraphics{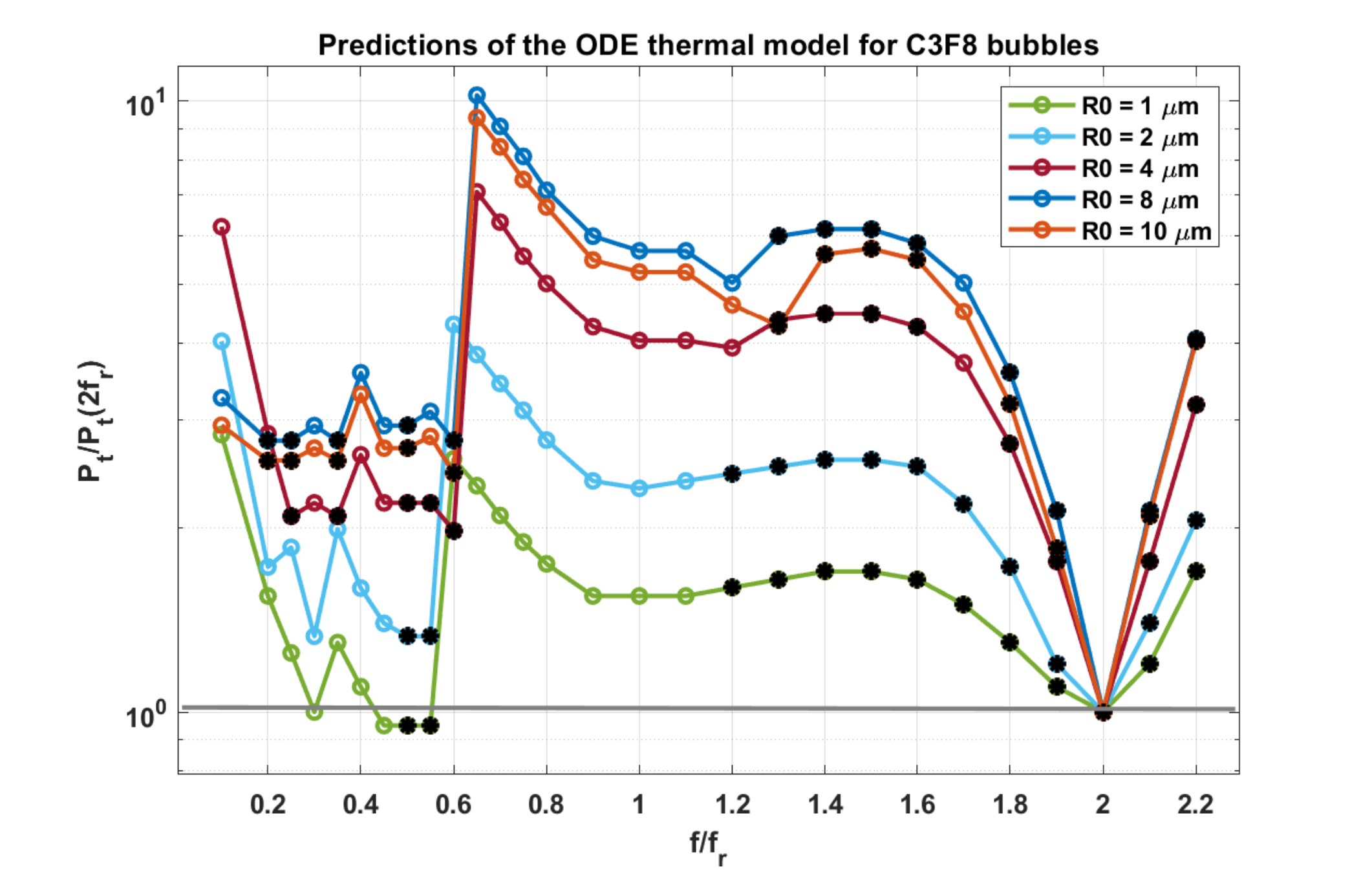}} \scalebox{0.3}{\includegraphics{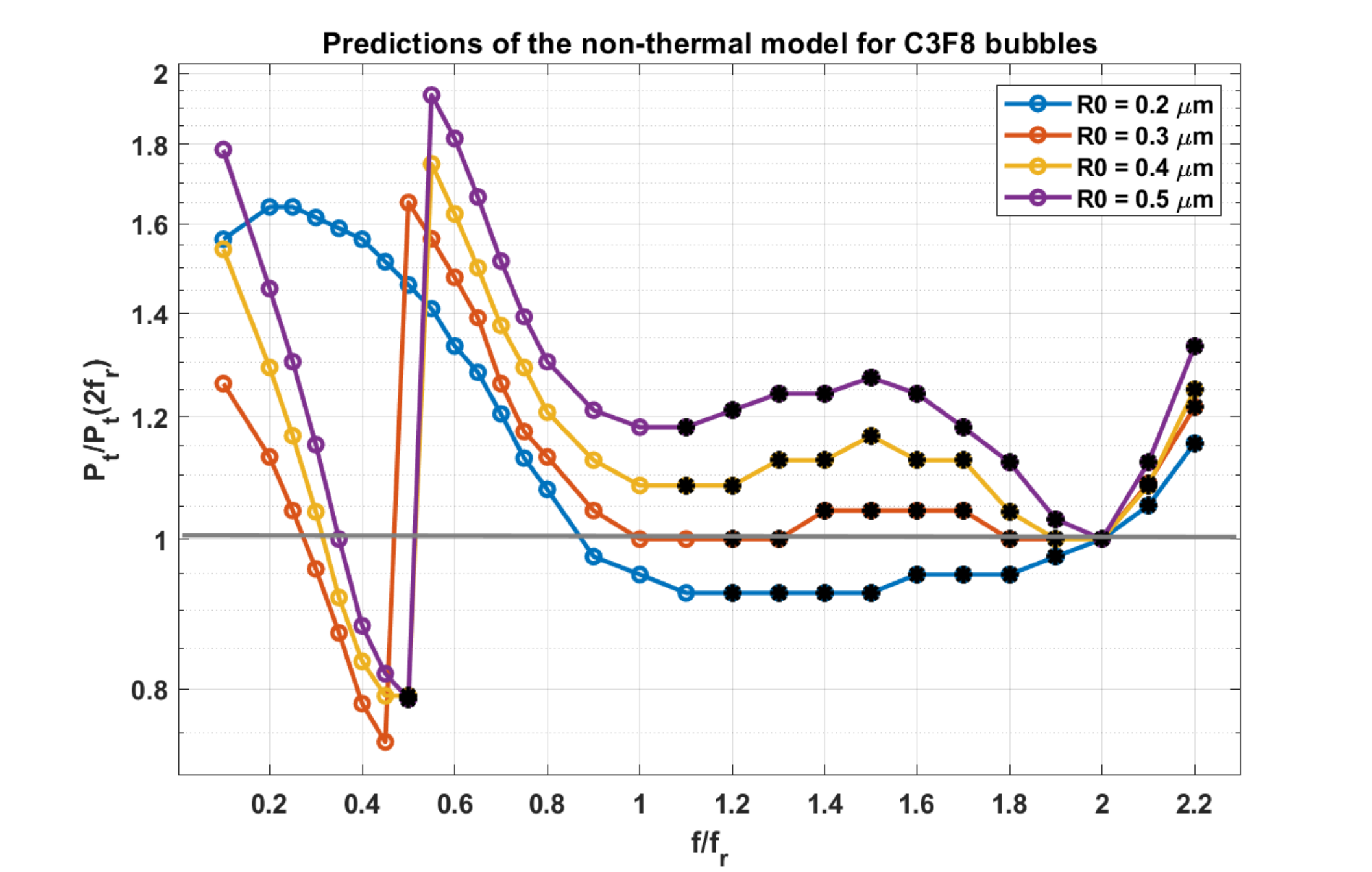}}\\
		\hspace{0.5cm} (e) \hspace{6cm} (f)\\
	\caption{Normalized pressure thresholds of the P2 oscillations of a C3F8 bubble as a function of frequency. The left column is for bubbles with initial radii between $1 \mu m\leq R_0 \leq 10 \mu m$ and the right column is for bubbles with initial radii between $ 0.2 \mu m\leq R_0 \leq 0.5 \mu m$: (a) $\&$ (b)- Non-thermal model, (c) $\&$ (d)-linear thermal (a) $\&$ (f)- ODE thermal model.  Solid black circles mark the stable oscillation regimes.}
	\end{center}
\end{figure*}
 Fig. 4 shows the pressure threshold of SH emissions as a function of frequency for C3F8 bubbles. In this case thermal damping is not as significant as air bubbles. For bubbles with $1 \mu m\leq R_0 \leq 10\mu m$ all three models of NTM (Fig. 4e), LTM (Fig. 4c) and FTM (Fig. 4e) predict the minimum at $f=2f_r$. The only exception is in case of the bubble with $R_0=1 \mu m$ where the FTM model predicts the minimum at $0.5f_r\leq f\leq 0.55f_r$. For C3F8  bubbles with $1 \mu m\leq R_0 \leq 10\mu m$, SH emissions are nondestructive for $0.5 f_r\leq f\leq 0.6f_r$ and $f\geq 1.2f_r$. The stability region can be extended to lower frequencies for larger bubbles. \\ For bubbles with $0.2 \mu m\leq R_0 \leq 0.5\mu m$, the NTM (Fig. 4b) and FTM (Fig. 4f) predict approximately the same behavior. The minimum is located at $f=0.5f_r$ for the bubbles with $0.3\mu m\leq R_0\leq 0.5 \mu m$. At this frequency FTM predicts stable SH emissions only for the bubble with $R_0=0.5 \mu m$. All three models predict stable SH emissions for $f\geq1.2f_r$.\\ Predictions of the LTM (Fig. 4d) deviate from the predictions of the FTM (Fig. 4f). This is because of the overestimation of the Td byb the LTM for smaller bubbles. The large dissipation as predicted by the LTM suppresses the UH resonance for bubbles $R_0\leq 0.4 \mu m$.     
\section{Limitations of the study}
In this study for highlighting the thermal dissipation effects we only considered spherical oscillations of a single bubble and one set of initial conditions ($R(t=0)=R_0$, $\dot{R}(t=0)=0 m/s$). However, non-spherical oscillations, especially in larger bubbles can significantly influence the bubble behavior \cite{4,63,64,65} and threshold of SH emissions \cite{31,32}. Bubble-bubble interaction is another important factor that changes the threshold of SH emissions in mono-disperse bubble clouds \cite{40} as well as poly-disperse bubble clouds \cite{66,67,68}. Interaction with a boundary also influences the pressure threshold of SH emissions \cite{68}. In real applications and especially for high bubble concentrations, the effect of bubble-bubble interactions should be considered for accurate predictions of the stable regions of SH emissions. The initial condition of the bubble is another factor that can change the minimum pressure threshold for the occurrence of nonlinear behavior and SH emissions \cite{3,69,70}. Sonication with a multi-frequency acoustic source \cite{7,71,72,73,74,75} significantly influences the nonlinear behavior of the bubbles and the threshold for SH emissions. For sufficiently long exposures, mass transfer effects become an important mechanism of the damping and can significantly affect the bubble behavior \cite{76,77,78,79,80}. Incorporation of these effects are beyond the scope of present study.  For accurate modeling of the cavitation phenomenon these effects must be included in the future studies. 
\section{conclusion}
In this study we demonstrated that the thermal damping significantly influences the pressure threshold of P2 oscillations and 1/2 order SH emissions. For gases with higher thermal damping like air, full thermal effects should be incorporated in the model to accurately predict the nonlinear oscillations. Predictions of the the popular linear thermal model \cite{41,42,49,50,81} significantly deviate from the predictions of the full thermal model that more accurately incorporates all the thermal effects. When thermal damping is considered with its full non-linearity, we show that the minimum pressure for period doubling and 1/2 order SH emissions for bubbles with large thermal damping (e.g. air) occurs when $0.5f_r \leq f\leq0.6f_r$  ($f_r$ is the linear resonance frequency). The mechanism of the period doubling is via the occurrence of 5/2  UH resonance. Results show that P2 oscillations and 1/2 order SH emissions are stable when $f\geq1.2f_r$ for all the studied bubble sizes ($0.2 \mu m\leq R_0\leq 10 \mu m$). However, only bubbles with $R_0\geq 0.5 \mu m$ ($\geq 1 \mu m$ in diameter) may exhibit stable SH emissions when $0.5f_r\leq f\leq 0.6f_r$. \\
\section{Acknowledgments}
	The work is supported by the Natural Sciences and Engineering Research Council of Canada (Discovery Grant RGPIN-2017-06496), NSERC and the Canadian Institutes of Health Research ( Collaborative Health Research Projects ) and the Terry Fox New Frontiers Program Project Grant in Ultrasound and MRI for Cancer Therapy (project $\#$1034). A. J. Sojahrood is supported by a CIHR Vanier Scholarship.

\end{document}